\documentclass[a4paper,11pt]{article}
\usepackage{amsmath}
\usepackage{amssymb}
\usepackage{amsthm}
\usepackage{stmaryrd}

\usepackage{graphicx,psfrag}%\usepackage{hyperref}
\usepackage[small]{caption}
\usepackage{subfig}

\usepackage{varioref} %NB: FIGURE LABELS MUST ALWAYS COME DIRECTLY AFTER CAPTION!!!
\vrefwarning

\usepackage{comment}

\graphicspath{{./images/}}

\usepackage[utf8]{inputenc}

\numberwithin{equation}{section}
\numberwithin{figure}{section}
\numberwithin{table}{section}
\usepackage{afterpage}%fingers crossed (The most important package)

\usepackage{array} 

\usepackage{multirow}

\usepackage{enumitem}

\newtheorem{thm}{Theorem}[section]

\newtheorem{lem}[thm]{Lemma}

\newcommand{\mbb}{\mathbb}

\renewcommand{\vec}[1]{\mathbf{#1}}

\newcommand{\rhat}{\hat{r}}

\newcommand{\vsqr}{V^2}

\newcommand{\vflat}{V_0}
\newcommand{\vsqflat}{\vsqr_0}

\newcommand{\inthinf}{\int_{0}^{\infty}}

\newcommand{\brkt}[1]{\left( #1 \right)}

\newcommand{\mudisc}{\mu}
\newcommand{\Mdisc}{M}

\newcommand{\Mdisctot}{M^{\text{gal}}}
\newcommand{\Mtot}{M^{\text{gal}}}

\newcommand{\divfrac}[2]{#1 / #2}

\newcommand{\rLmin}{a_{<}}
\newcommand{\rRmax}{a_{>}}

\renewcommand{\epsilon}{\varepsilon}

\usepackage{jcappub} % for details on the use of the package, please
\newcommand{\dz}{\cdot}
\newcommand{\Lcal}{\mathcal{L}}
\newcommand{\Rcal}{\mathcal{R}}

\graphicspath{{./images/}}

\title{\textbf{Galactic Disc Rotation: Analytic models and asymptotic results.}}

\author{Niall Ryan}
\affiliation{University of Limerick,\\Ireland}

\emailAdd{niall.ryan@ul.ie}

\abstract{This paper outlines an exact analytic model for self-gravitating thin disc galaxies with flat rotation curves. It is shown that thin discs of matter alone can support perfectly flat rotation curves under Newtonian gravity, without needing extended halos or modified gravity. The method allows disc density to be inferred from polynomial approximations to the rotation curve. Several expository models are presented. The model also shows the link between flat rotation curves and ``exponential'' discs. It is also shown that reconstructed discs inherently posses vast mass ``hinterlands'' in their outer regions, that can contain the majority of their total mass. Some implications for the problem of missing mass are discussed.}

\begin{document}

% \title{\textbf{Galactic Disc Rotation: Analytic models and asymptotic results.}}
% \author{Niall Ryan\thanks{E-mail:
% niall.ryan@ul.ie}\\
% \emph{Dept. of Mathematics, University of Limerick, Limerick, Ireland}}

%\email[Email: ]{niall.ryan@ul.ie}
%\homepage[]{Your web page}
%\thanks{}
%\altaffiliation{}
%\affiliation{Dept. of Mathematics, University of Limerick, Limerick, Ireland}

\date{\today}
\maketitle
%\flushbottom

\begin{abstract}

\end{abstract}

%\maketitle

% \begin{keywords}
% gravitation
% celestial mechanics
% galaxies: structure
% galaxies: kinematics and dynamics
% \end{keywords}

% \begin{multicols}{2}
% \footnotesize
% \tableofcontents
% \end{multicols}

\section{Introduction}

It is often assumed that flat rotation curves in spiral galaxies cannot be explained by discs of baryonic matter acting under Newtonian gravity. Such discs are instead expected to produce Keplerian velocity fall-offs. However, this is not the case, and in fact discs alone can support flat rotation curves of arbitrary length without needing spherical halos or modified gravity.

\begin{figure}%[htbp]
  \centering
  \includegraphics[width=0.49\textwidth]{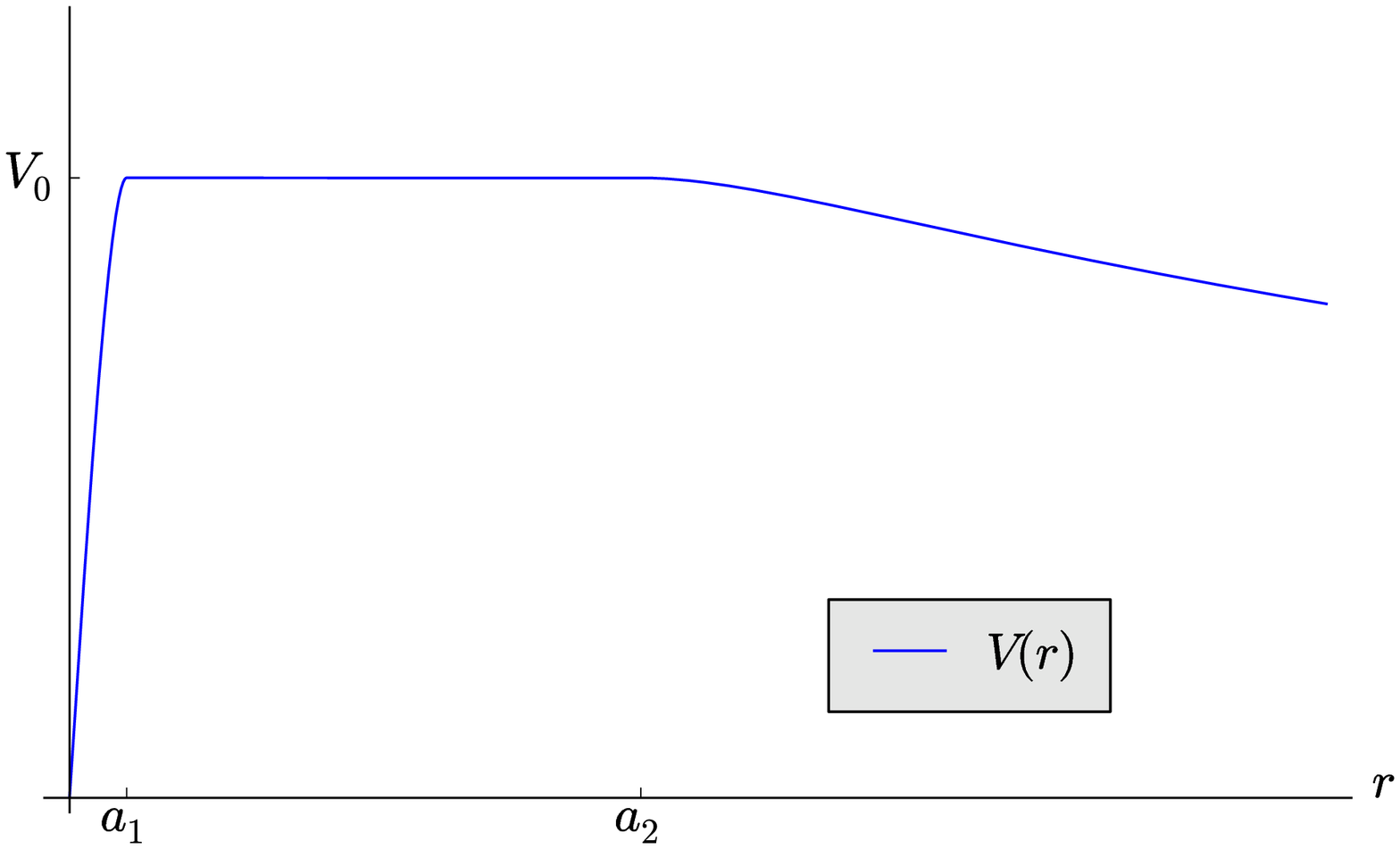}
  \includegraphics[width=0.49\textwidth]{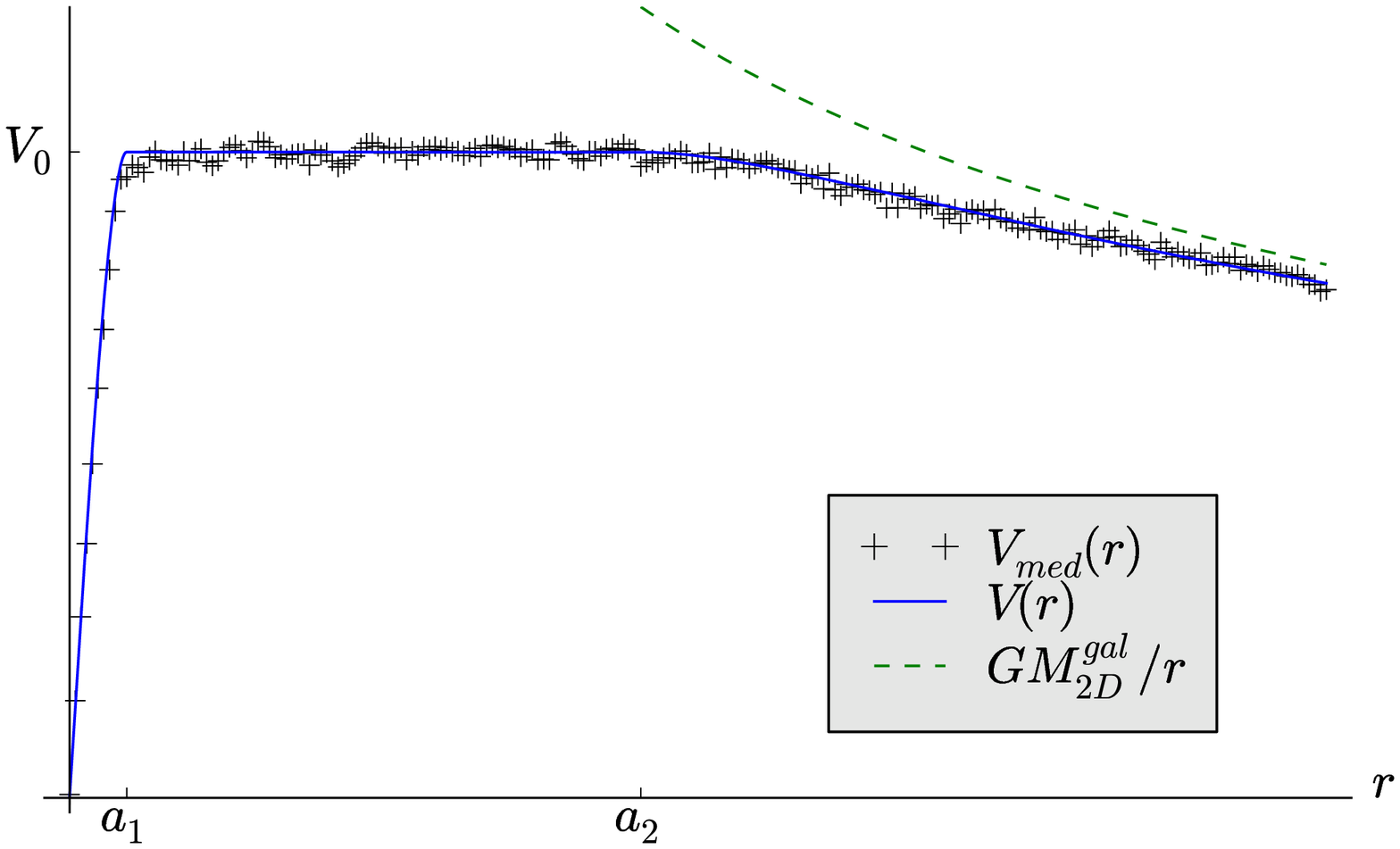}
  \caption{\textbf{Left:} A rotation curve $V(r)$ of a self gravitating galactic disc, of the form (\ref{eq:vp_piecewise}), which is is constant or ``flat'' between the distances $r=a_1$ and $r=a_2$.\\
\textbf{Right:} The rotation curve $V_{\text{med}}$ ($+$) of a simulated planar disc of $10^6$ masses, generated to support $\vsqr(r)$ (see section \ref{sec:galaxies-gas-nebulae}, Figure \ref{fig:gal_plot_log_10}). The rotation curve of a point mass at $r=0$ with mass equal to that of the entire disc is also shown.
}
  \label{fig:P_VEL_MODEL_PLOT}
\end{figure}

Let the square of the rotation curve, $\vsqr(r)$, be a axisymmetric velocity profile of the form
\begin{equation}
\label{eq:vp_piecewise}
 \vsqr(r) = \begin{cases}
\vsqr_L(r), & r<a_1 \\
\vsqr_M(r)= \vsqr_0, & a_1<r<a_2 \\
\vsqr_R(r), & a_2<r
\end{cases}  
\end{equation}
where $r$ is distance from the axis of rotation, and $a_1 < a_2$ are characteristic length scales. Here it will be assumed that $\vsqr_L(0)=0$, and that $\vsqr_M(r) = \vsqflat$ is constant or ``flat'' between $a_1$ and $a_2$. Such a profile is plotted in Figure \ref{fig:P_VEL_MODEL_PLOT}(Left). The goal is to find thin discs capable of supporting such velocity profiles.

Velocity profiles are typically matched to galaxy mass distributions in one of two ways\cite{Sofue2013}. Firstly, numerical methods can be used to directly calculate mass density distributions from observed data\cite{Nordsieck1973,Clutton-Brock1972,Kostov2007}. While these methods give mass distributions using minimal prior assumptions, they can be computationally intensive and the results opaque to analysis.

Alternatively, so called decomposition models can be used, in which the galaxy is divided into discrete components -- disc, bulge, halo, etc -- and the total velocity profile is assumed to be the sum of contributions from these components with,
\[ \vsqr = \vsqr_{\text{disc}} +\vsqr_{\text{bulge}}+ \vsqr_{\text{halo}}\]
A model for each component is chosen and the masses, scales, and other parameters in each are adjusted to reproduce the desired $\vsqr$. Using this technique galaxy models can be fitted to observed rotation curves, and the resulting mass distributions are moreover analytical \cite{Palunas2000,Bosma1981}.

Several models exist for each component type. For bulges there are point masses, homogeneous and isothermal spheres, Plummer-Kuzmin, isochrone, and many other models \cite{Binney_Tremaine_GalDy}. Some of these are also used for halos, along with more specialised models such as the NFW profile\cite{Navarro1996}. There is also a wide selection of models for discs including exponential discs\cite{Freeman1970}, Kuzmin-Toomre models\cite{kuzmin1952,Toomre1964}, power law discs\cite{Sellwood2001}, and Mestel's Disc\cite{1963MNRAS.126..553M}. There are also hybrid bulge/disc models such as those of Miyamoto and Nagai\cite{1975PASJ...27..533M}.

It should be noted that most of these component models have relatively few free parameters -- typically including mass and scale. This makes the models analytically simple but practically rigid, as their forms are more or less fixed. The flexibility to fit observed rotation curves only comes from combining multiple components.

In particular, none of the component models are flexible enough to reproduce flat rotation curves by themselves -- the exceptions being isothermal spheres and Mestel's Disc\footnote{However these models result in galaxies with infinite densities and infinite total mass respectively.}. Since the phenomenon of flat rotation curves is commonly seen in spiral galaxies\cite{Rubin1970,Rubin1980,Sofue2001} decomposition models must therefore be frequently employed with most of these needing an extended dark matter halo to explain extended flat rotation curves.

However, the combination of components required to reproduce observed curves gives rise to questions of its own. Among these are the problems of the disc-halo conspiracy, maximum disc fitting, as well as the nature and geometry of the halos themselves. These and several other criticisms have been outlined by Davies\cite{Davies2012}. Others have proposed alternative hypotheses, such as MOND\cite{Milgrom1983}, which use modified dynamics or gravity to explain flat rotation curves. These too have raised criticisms\cite{2004IAUS..220...27S}.

In the author's opinion, the present impasse in understanding flat rotation curves stems from the rigidity of existing disc models, which force the use of halos or modified gravity at large distances. Therefore, this paper will introduce a highly flexible thin-disc model, capable of fitting very general rotation curves, including those with sections that are absolutely flat, and without needing separate galaxy components. The price of such flexibility is a certain degree of analytic complexity, but which ultimately amounts to the evaluation of convergent power series that can be easily and speedily done with modern computers.

For the sake of simplicity in what follows, only thin planar disc models will be considered -- without bulges or halos. Moreover, the unrealistic but simplifying assumptions of collision-less, circular orbits will be used. Questions of stability will also not be addressed. The model presented here is intended to be a first step towards an alternative disc centered framework for understanding flat rotation curves, 

One restriction made is that only physically reasonable discs with finite total mass and everywhere positive density will be considered in this paper. It will be seen that this restriction constrains rotation curves to be asymptotically Keplerian at infinity. For this reason, rotation curves which remain flat indefinitely, such as Mestel's Disc, will not be considered in this paper.

\section{Toomre's disc galaxy model}
Consider an infinitely thin, axisymmetric, planar disc galaxy with surface density profile $\mu(r)$. Assuming that the material of the disc is in collision-less, self-gravitationally bound circular orbit with orbital speed(rotation curve) $V(r)$, Toomre \cite{1963ApJ...138..385T}(Eq. (12)) has shown that $\mu(r)$ can be derived from $\vsqr(r)$ using the doubly infinite integral
\begin{equation}
  \label{eq:1}
  2 \pi G \mu(r) = \int_{0}^{\infty } \frac{d \vsqr(\rhat)}{d \rhat}\int_{0}^{\infty } J_0 \left( k \rhat \right) J_0 \left( k r \right)  dk d\rhat
\end{equation}
Where $J_0$ is Bessel's function of order $0$ (See \cite{G+R} $\S 8.4$). Note that the integral is linear in the \emph{velocity profile} $\vsqr$, rather than $V$, and so it is $\vsqr$ that is focused on in what follows. Equation \eqref{eq:1} is a formal solution to the inverse problem\cite{inverse_probs_activ} of determining a disc's density from its rotation curve.

Various special cases and families of analytic velocity/density profile pairs $(\vsqr,\mu)$ for self gravitating discs have been found using Toomre's method\cite{1963ApJ...138..385T,Freeman1970,Clutton-Brock1972}, including an infinite family of models and a Gaussian profile introduced by Toomre in the same paper. Pairs satisfying \eqref{eq:1} have also been found by independent means \cite{1963MNRAS.126..553M,1992MNRAS.257..581Q,Kalnajs1971}, the most notable of which is Mestel's Disc\cite{1963MNRAS.126..553M}, defined by the profile pair $\vsqr(r)=\vsqr_0$, $2\pi G \mu(r) = \vsqr_0/r$, which has a rotation curve that is \emph{everywhere flat}, but whose resulting mass is infinite.

However, as noted above, such analytic models are specialised, and can be difficult to fit to observed rotation curve data. The models can be augmented using a decomposition as mentioned above, which introduces additional galaxy components such as a bulge or halo. Alternatively direct numerical methods can be used to solve \eqref{eq:1} directly using discs alone, but such procedures are complicated by the difficulty of the integral and by the need to extrapolate $\vsqr(r)$ in outer regions\cite{2011RAA....11.1429F,Shatskiy2012,Nordsieck1973}.

A key aim of this paper is to find an analytic framework for profile pairs ($\vsqr$,$\mu$) which is capable of describing a wide range of observed velocity profiles, including ones such as \eqref{eq:vp_piecewise} which have regions of absolutely flat velocity. This is accomplished by using polynomial approximations to $\vsqr(r)$ in (\ref{eq:1}), and by representing the integral's kernel using infinite power series. This allows very general and versatile disc models to be constructed quickly, without the need for numerical integration, and in a way that produces a fully analytic disc density profile which is open to general mathematical analysis.

\subsection{A series for Toomre's integral}

Equation (\ref{eq:1}) is a robustly improper integral, and its evaluation is the principal difficulty with using Toomre's method. The integral is extremely sensitive to both the local and asymptotic behaviour of $\vsqr(r)$. To make (\ref{eq:1}) more amenable to analysis, following Toomre the kernel over $k$ can be evaluated using formulas $\S6.512(1)$ and $\S 8.113(1)$ of Gradshteyn and Ryzhik \cite{G+R}. For $r \ne \rhat$ these give, 
\begin{equation}
  \label{eq:2}
\int_{0}^{\infty } J_0 \left( k \rhat \right) J_0 \left( kr \right) dk= \frac{2}{\pi r_>} K\left(\frac{r_<}{r_>}\right)
\end{equation}
where $K(k)$ is the complete elliptic integral of the first kind(see appendix \ref{sec:preliminary-lemmas}), and where
\begin{equation}
  \label{eq:rmin_rmax_shorthands_def}
 r_<=\min(r,\rhat)\quad  \text{and}\quad  r_>=\max(r,\rhat)
\end{equation}
are shorthands for the smaller and larger of the lengths $r$ and $\rhat$. Now, $K$ has Maclaurin series $\left(2/\pi\right)K(k) = \sum_{m=0}^{\infty } G_{2m} k^{2m}$ (see \cite{G+R} $\S 8.113(1)$), where the coefficients $G_{2m}$ are given by either of the representations\footnote{Here the first representation uses the double factorial ($!!$)}
\begin{equation}
  \label{eq:3}
  G_{2m} = \left[ \frac{ \left( 2m-1 \right)!! }{ \left( 2m \right)!!} \right]^2 =  \left[ \frac{(2m)!}{2^{2m} (m!)^2}\right]^2
\end{equation}
Using this series in (\ref{eq:2}), equation (\ref{eq:1}) can be rewritten as
\begin{equation}
  \label{eq:4}
  2 \pi G \mu (r)= \sum_{m=0}^{\infty} G_{2m} \inthinf \frac{d \vsqr(\rhat)}{d \rhat} \frac{r_<^{2m}}{r_>^{2m+1}} d \rhat
\end{equation}
And so $\mu(r)$ has been represented as an infinite series of integrals over $\rhat$ only. This result facilitates straightforward symbolic integration, particularly when the velocity profile $\vsqr$ is piecewise polynomial.

\section{Disc model framework}\label{sec:disc-model-framework}

With $\vsqr$ of the form (\ref{eq:vp_piecewise}), splitting the integral in (\ref{eq:4}) and applying $\divfrac{d \vsqr_M}{d r}=0$ gives
\begin{align} 2 \pi G \mudisc (r) &=\sum_{m=0}^{\infty} G_{2m} \left(\int_0^{a_1} + \int_{a_1}^{a_2} + \int_{a_2}^{\infty}\right)  \frac{d \vsqr(\rhat)}{d \rhat} \frac{r_<^{2m}}{r_>^{2m+1}} d \rhat\\
&=2\pi G\mu_L(r) + 2 \pi G\mu_R(r)
\end{align}
where
\begin{align}
  \label{eq:rho_L_def} 2\pi G\mu_L(r)&=\sum_{m=0}^{\infty} G_{2m} \int_0^{a_1} \frac{d \vsqr_L(\rhat)}{d \rhat} \frac{r_<^{2m}}{r_>^{2m+1}} d \rhat\\
\label{eq:rho_R_def} 2\pi G \mu_R(r)&=\sum_{m=0}^{\infty} G_{2m} \int_{a_2}^{\infty} \frac{d \vsqr_R(\rhat)}{d \rhat} \frac{r_<^{2m}}{r_>^{2m+1}} d \rhat
\end{align}
The terms $\mu_L(r)$ and $\mu_R(r)$ will be referred to as the left and right density profile parts respectively. These density profile parts are the contributions to $\mudisc$ from the left and right velocity profile parts $\vsqr_L$ and $\vsqr_R$ respectively. Analysing these functions separately allows for $\vsqr_L$ and $\vsqr_R$ to be chosen independently of one another.

\subsection{Elementary density profiles}
With an eye to representing the velocity profile parts $\vsqr_L$ and $\vsqr_R$ using polynomial series, consider very simple monomial profile parts of the form
\begin{equation}
\label{eq:monomial_v2LR}
\vsqr_L(r)=\vsqflat (r/a_1)^n,\quad \text{ and} \quad  \vsqr_R(r)=\vsqflat (a_2/r)^n,
\end{equation}
for general indices $n>0$, and constant flat velocity $\vflat=\vsqr_M$. Then, let $\mu_L(r;n)$ and $\mu_R(r;n)$ denote the corresponding left and right \emph{elementary radial density profiles}, given by \eqref{eq:rho_L_def}/\eqref{eq:rho_R_def}. Before stating these functions, first an important sequence of constants must be defined.

\begin{table}%[t]
    \centering
    \begin{tabular}{|c|c|}
      \hline
      $n$ & $W_n$\\
      \hline
        $0$ & $2 \, \ln\left(2\right)$ \\
        $1$ & $0$ \\  
        $2$ & $\frac{1}{2} \, \ln\left(2\right) - \frac{1}{4}$ \\
 
        $3$ & $0$ \\  
        $4$ & $\frac{9}{32} \, \ln\left(2\right) - \frac{21}{128}$
\\  
        $5$ & $0$ \\  
        $6$ & $\frac{25}{128} \, \ln\left(2\right) -
\frac{185}{1536}$ \\  
        $7$ & $0$ \\ \hline 
    \end{tabular}
    \caption{Values of the constants $W_n$ for small $n$. See equation (\ref{eq:w_n_log_def})}
    \label{tab:w_n_table_text} 
\end{table}

Let the sequence of constants $W_n$ be defined for even and odd $n$ by the formula
\begin{equation}
\label{eq:Wn_Def_text_sum}
 W_n = \sum_{m=0}^\infty \frac{G_{2m}}{2m+n+1} + \sum_{\substack{m=0\vspace{0.1em}  \\2m\ne n} }^\infty \frac{G_{2m}}{2m-n} 
\end{equation}
These constants are ubiquitous in the analysis of series arising from (\ref{eq:4}). The first few values of $W_n$ are given in Table \ref{tab:w_n_table_text}, and an equivalent form is given by equation (\ref{eq:w_n_log_def}). In particular, note that $W_1=0$.

Recall the sequence $G_n$ from (\ref{eq:3}), and let the binary indicator sequence $B_n$ be $B_n=1$ when $n$ is even, and $B_n=0$ when $n$ is odd. Then using the constants $W_n$, piecewise series for the elementary density profiles $\mu_L(r;n)$ and $\mu_R(r;n)$ may now be stated in the following lemmas, whose complete proofs are outlined in Appendix \ref{sect:app_elem_den_results}.
%\begin{widetext}
\begin{lem}\label{lem:muL_series}
  If $\vsqr_L(r)=\vsqflat (r/a_1)^n$ for $n>0$, then the corresponding $\mu_L$ defined by \eqref{eq:rho_L_def} is given by
  \begin{equation}
    \label{eq:rho_L_piecewise_text}
    2 \pi G \mu_L(r;n)= \begin{cases}
      2 \pi G \mu_{LL}(r;n) &, r< a_1\\
      2 \pi G \mu_{LR}(r;n) &, r> a_1
    \end{cases}
  \end{equation}
  where $\mu_{LL}$ and $\mu_{LR}$ are the left and right parts of $\mu_L$, that are given by the series
  \begin{align}
    \label{eq:rho_LL_text} 2 \pi G  \mu_{LL}(r;n)
     &=\vsqflat \frac{n}{a_{1}}\Bigg[\sum_{\begin{subarray}{c} m=0\\ 2m \ne n-1\end{subarray}}^{\infty}\frac{-G_{2m}}{2m-n+1}\left(\frac{r}{a_{1}}\right)^{2m}
+\bigg[W_{n\text{-}1}-B_{n\text{-}1}G_{n\text{-}1}\ln\left(\frac{r}{a_{1}}\right)\bigg]\left(\frac{r}{a_{1}}\right)^{n-1}\Bigg]\\
    \label{eq:rho_LR_text} 2 \pi G  \mu_{LR}(r;n)&= \vsqflat\frac{n}{a_{1}}\sum_{m=0}^{\infty}\frac{G_{2m}}{2m+n}\left(\frac{a_{1}}{r}\right)^{2m+1}
  \end{align}
\end{lem}
\begin{lem}\label{lem:muR_series}
  If $\vsqr_R(r)=\vsqflat (a_2/r)^n$ for $n>0$, then the corresponding $\mu_R$ defined by \eqref{eq:rho_R_def} is given by
  \begin{equation}
    \label{eq:rho_R_piecewise_text}
    2 \pi G \mu_R(r;n)= \begin{cases}
      2 \pi G \mu_{RL}(r;n) &, r< a_2\\
      2 \pi G \mu_{RR}(r;n) &, r> a_2
    \end{cases}
  \end{equation}
  where $\mu_{RL}$ and $\mu_{RR}$ are the left and right parts of $\mu_R$, that are given by the series
  \begin{align}
    \label{eq:rho_RL_text} 2 \pi  G \mu_{RL}(r;n)&=-\vsqflat \frac{n}{a_{2}}\sum_{m=0}^{\infty}\frac{G_{2m}}{2m+n+1}\left(\frac{r}{a_{2}}\right)^{2m}\\
    \label{eq:rho_RR_text} 2 \pi  G  \mu_{RR}(r;n) & =-\vsqflat \frac{n}{a_{2}}\Bigg[-\sum_{\begin{subarray}{c} m=0\\ 2m\ne n\end{subarray}}^{\infty}\frac{G_{2m}}{2m-n}\left(\frac{a_{2}}{r}\right)^{2m+1}  +\bigg[W_{n}-B_{n}G_{n}\ln\left(\frac{a_{2}}{r}\right)\bigg]\left(\frac{a_{2}}{r}\right)^{n+1}\Bigg]
  \end{align}
\end{lem}
Though lengthy, these definitions amount to straightforward piecewise power series for $\mu_L(r;n)$ and $\mu_R(r;n)$. The change in definitions results from various integrals vanishing or appearing, due to the changes in $r_<$ and $r_>$ at $r=a_1$ and $r=a_2$. Moreover though not proven here, it is the case that both $\mu_L(r;n)$ and $\mu_R(r;n)$ and their derivatives $\mu_L'(r;n)$ and $\mu_R'(r;n)$ are continuous for all $r>0$, including across $r=a_1$ and $r=a_2$. Thus for all $n>0$, $\mu_L(r;n)$ and $\mu_R(r;n)$ are smooth, continuous functions, given by elementary power series in $r/a_1$ and $r/a_2$. 

\subsection{Polynomial velocity profiles}\label{sec:polyn-veloc-prof}
Now, these elementary density profile parts correspond to the monomial velocity profiles in  \eqref{eq:monomial_v2LR}. But by the linearity of \eqref{eq:rho_L_def}, \eqref{eq:rho_R_def}, and (\ref{eq:4}) in $\vsqr$, if the left and right velocity profiles are given as finite\footnote{Though important, this paper does not consider infinite series for $\vsqr(r)$ and whether the corresponding density profile series in such cases are convergent.} polynomial series of the forms
\begin{align}
%\notag  \vsqr_L(r)&=\vsqflat  \sum_n \Lcal_n \brkt{\frac{r}{a_1}}^n\\
% \label{eq:polynomial_vel_profile_parts}  \text{and} \quad  \vsqr_R(r)&=\vsqflat  \sum_n \Rcal_n \brkt{\frac{a_2}{r}}^n 
 \label{eq:polynomial_vel_profile_parts} \vsqr_L(r)=\vsqflat  \sum_n \Lcal_n \brkt{\frac{r}{a_1}}^n  \quad \text{and} \quad  \vsqr_R(r)&=\vsqflat  \sum_n \Rcal_n \brkt{\frac{a_2}{r}}^n 
\end{align}
then the corresponding density profile parts are then given by the series
\begin{align*}
%  \mu_L(r)&= \sum_n \Lcal_n\, \mu_L(r;n)\\
%\text{and} \quad  \mu_R(r)&=\sum_n \Rcal_n\, \mu_R(r;n).
\mu_L(r)= \sum_n \Lcal_n\, \mu_L(r;n) \quad \text{and} \quad  \mu_R(r)=\sum_n \Rcal_n\, \mu_R(r;n).
\end{align*}
Here $\Lcal_n$ and $\Rcal_n$ are dimensionless polynomial coefficients. From these the total radial density is given by
\begin{equation}
  \label{eq:mudisc_constants_sum}
  2 \pi G\mudisc(r) = 2 \pi G \sum_n \Lcal_n \mu_L(r;n)+ \Rcal_n \mu_R(r;n)
\end{equation}
Expanding this piecewise using \eqref{eq:rho_L_piecewise_text} and \eqref{eq:rho_R_piecewise_text}, and using \eqref{eq:polynomial_vel_profile_parts} in (\ref{eq:vp_piecewise}), it can thus be concluded that a partially flat velocity profile of the form
\begin{equation}
\label{eq:vp_piecewise_polynomial} 
\vsqr(r) = \vsqflat \times \begin{cases}
\sum_n \Lcal_n \brkt{\divfrac{r}{a_1}}^n , & r<a_1 \\
1 & a_1<r<a_2 \\
\sum_n \Rcal_n \brkt{\divfrac{a_2}{r}}^n, & a_2<r
\end{cases}
\end{equation}
is supported under Newtonian gravity by a thin disc with radial surface density profile
\begin{align} 
\label{eq:rhodisc_polynomial} 2 \pi G &\mudisc(r)=2 \pi G \, \times \begin{cases}
\sum_n \Lcal_n \mu_{LL}(r;n)+\Rcal_n \mu_{RL}(r;n), & r<a_1 \\
\sum_n \Lcal_n \mu_{LR}(r;n)+\Rcal_n \mu_{RL}(r;n), & a_1 < r<a_2 \\
\sum_n \Lcal_n \mu_{LR}(r;n)+\Rcal_n \mu_{RR}(r;n), & a_2<r \\
\end{cases}
\end{align}
where the functions $\mu_{LL}$, $\mu _{LR}$, $\mu _{RL}$, and $\mu _{RR}$ are those given in lemmas \ref{lem:muL_series} and \ref{lem:muR_series}.

The piecewise formulas \eqref{eq:vp_piecewise_polynomial} and \eqref{eq:rhodisc_polynomial} constitute an analytic velocity/density profile pair $(\vsqr,\mu)$ satisfying (\ref{eq:1}). Determining $\mu(r)$ from any given piecewise polynomial $\vsqr(r)$ simply involves evaluating a finite number of elementary density profiles, with known convergent power series. In practice this method can be used to rapidly generate and analytically investigate a very wide class of self gravitating disc models.

In should be noted that this paper does not derive conditions which can guarantee everywhere strictly positive $\mu(r)$. Empirical investigations suggest that in general, more slowly varying velocity profiles typically give everywhere positive $\mu(r)$, and conversely that rapidly declining $\vsqr(r)$ lead to negative or very sparse densities. However, no firm conditions or limits have been found owing to the ill-posedness which is typical of such inverse problem(See \cite{inverse_probs_activ}). However, conditions guaranteeing finite total disc mass are derived later in section \ref{sec:asympt-restr-veloc}.

\subsection{Example models}\label{sec:example-models}

To demonstrate the utility of this method, several examples of such analytic disc galaxy models are now constructed.
\begin{table}%[b]
    \centering
%\begin{ruledtabular}
    \begin{tabular}{|c||c|c|c||c|c|c|c||c|}
\hline
Model& $\Lcal_2$& $\Lcal_3$& $\Lcal_4$& $\Rcal_1$& $\Rcal_2$& $\Rcal_3$& $\Rcal_4$& $\divfrac{a_1}{a_2}$\\
\hline
A& 1& $\dz$& $\dz$& 1& $\dz$& $\dz$& $\dz$& 1/10\\
%\hline
B& 2& $\dz$& -1& 3/2& $\dz$& -1/2& $\dz$& 1/10\\
%\hline
C&2& $\dz$& -1& 2& -1& $\dz$& $\dz$& 1/10\\
%\hline
D&2& $\dz$& -1& 4& -6& 4& -1& 1/10\\
%\hline
E&2& $\dz$& -1& 4& -6& 4& -1& 1/2\\
%\hline
F&3& -2& $\dz$& 3/2& $\dz$& -1/2& $\dz$& 1/3\\
%\hline
G&3& -2& $\dz$& 3/2& $\dz$& -1/2& $\dz$& 1/2\\
%\hline
H&3& -2& $\dz$& 3/2& $\dz$& -1/2& $\dz$& 1\\
%\hline
I&12& -20& 9& 6& -9& 4& $\dz$& 1/5\\
%\hline
J&18& -32& 15& 2& -6& 10& -5& 1/5\\
\hline
    \end{tabular}
%\end{ruledtabular}
    \caption{Ten flat disc models, based on their velocity polynomial coefficients $\Lcal_n$,$\Rcal_n$, and the ratio $a_1/a_2$. A cipher($\dz$) denotes a $0$ coefficient. All non-listed coefficients are also taken to be $0$. }
    \label{tab:various_disc_model_paramaters} 
\end{table}
Any model pair \eqref{eq:vp_piecewise_polynomial}/\eqref{eq:rhodisc_polynomial} is defined completely by its velocity profile coefficients $\Lcal_n$ and $\Rcal_n$, and by the characteristic parameters $a_1,a_2$ and $\vsqr_0$. Taking $\vsqr_0$ and outer length $a_2$ to be arbitrary but fixed constants, consider the models given in Table \ref{tab:various_disc_model_paramaters}. Each of the models A-J lists the values of its coefficients $\Lcal_2,\Lcal_3,\Lcal_4,\Rcal_1,\Rcal_2,\Rcal_2,\Rcal_4$, with all other coefficients $\Lcal_n,\Rcal_n$ taken to be zero. In addition, the ratios $a_1/a_2$ in each model are also listed.

The velocity profiles $\vsqr(r)$, and density profiles\footnote{The density profiles are plotted on a normalised log scale.} $\mu(r)$ for the ten models A-J are plotted in Figures \ref{fig:P_MODELS_VELS} and \ref{fig:P_MODELS_DENS} respectively; The velocity profiles of all models are continuous, with $\vsqr(0)=0$ and Keplerian $O(1/r)$ behaviour at infinity. It has also been checked that these models produce everywhere positive densities $\mu(r)$ and so represent physically valid discs.

The ``brightness profiles'' in Figure \ref{fig:P_MODELS_DENS} have been scaled relative to the central density $\mu_0=\mu(0)$ in each model, which can be calculated using (\ref{eq:rho_LL_text}), (\ref{eq:rho_RL_text}), and (\ref{eq:mudisc_constants_sum}) as
\begin{equation}
  \label{eq:rhodisc_zero}
  2 \pi G \mudisc(0) = \frac{\vsqr_0}{a_1} \sum_n \Lcal_n\frac{n}{n-1} - \frac{\vsqr_0}{a_2}  \sum_n \Rcal_n\frac{n}{n+1} 
\end{equation}

Several general observations can be made from these figures. Firstly as radius $r$ increases, density decreases quickly past $r=a_1$, with this rate of decrease continually slowing past $r=a_2$. The slow curling upwards of $\log(\mu) \sim -3 \log(r)$ for $r > a_2$ reflects the fact that density $\mu(r)$ has asymptotic $O(1/r^3)$ behaviour as $r \to \infty$ (See section \ref{sec:asympt-restr-veloc}).

Secondly, it can be seen that models with sharp changes in velocity -- models A,I and J -- exhibit correspondingly sharp changes in density in the same locations. Model A exhibits a kink in $\mu(r)$ at $r=a_2$, corresponding to the sudden transition from $\vsqr_M = \vsqr_0$ to $\vsqr_R = \vsqr_0 (a_2/r)$. Model J in particular has extreme localised changes in velocity and density, and note that its behaviour for $r>a_2$ results from velocity temporarily decreasing at a faster than Keplerian rate -- this can be seen in the region where the velocity profile for model J dips below that for model A in \ref{fig:P_MODELS_VELS}. Conversely, models with smooth velocity profiles have overall smoother density profiles.

Thirdly, note that in model H, $a_1=a_2$, and so this model has no flat velocity region. However the model framework \eqref{eq:vp_piecewise_polynomial}/\eqref{eq:rhodisc_polynomial} still applies, and does not require such a region to be present in order to work. Note also that model H exhibits no pseudo-exponential behaviour(see below), as this is tied to the existence of a flat velocity region.

\afterpage{
\begin{figure*}%[htbp]
  \centering
  \includegraphics[width=0.8\textwidth]{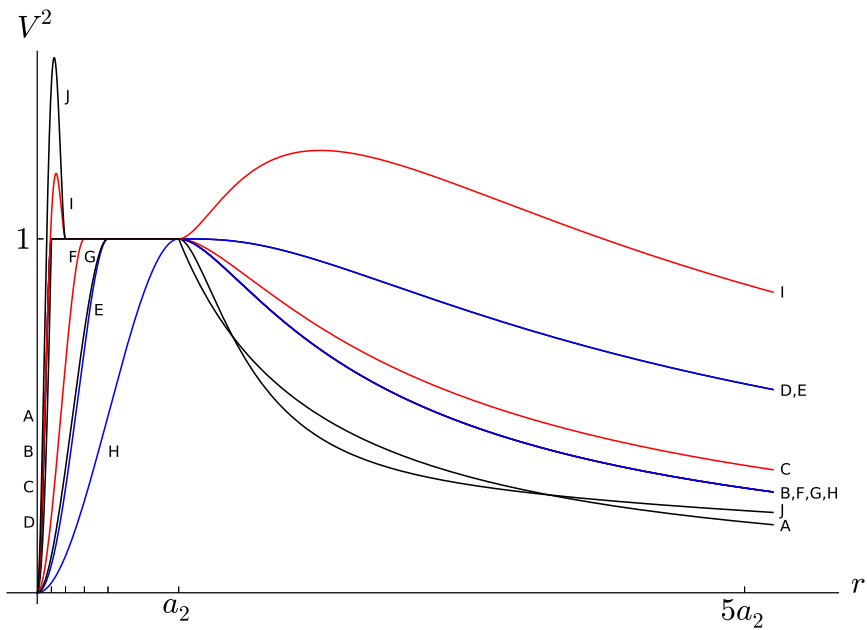}
\caption{The velocity profiles $\vsqr(r)$ of disc models A-J in Table \ref{tab:various_disc_model_paramaters}. Most models have smoothly changing velocities with flat regions. However, the profiles of model A,I and J exhibit sharp changes in velocity, and model H has no region of flat velocity. The length scale $a_2$ is arbitrary but fixed. The ratio $a_1/a_2$ varies across models, with the locations of $a_1$ marked on the $r$ axis.}
  \label{fig:P_MODELS_VELS}
\end{figure*}

 \begin{figure*}%[htbp]
  \centering
  \includegraphics[width=0.8\textwidth]{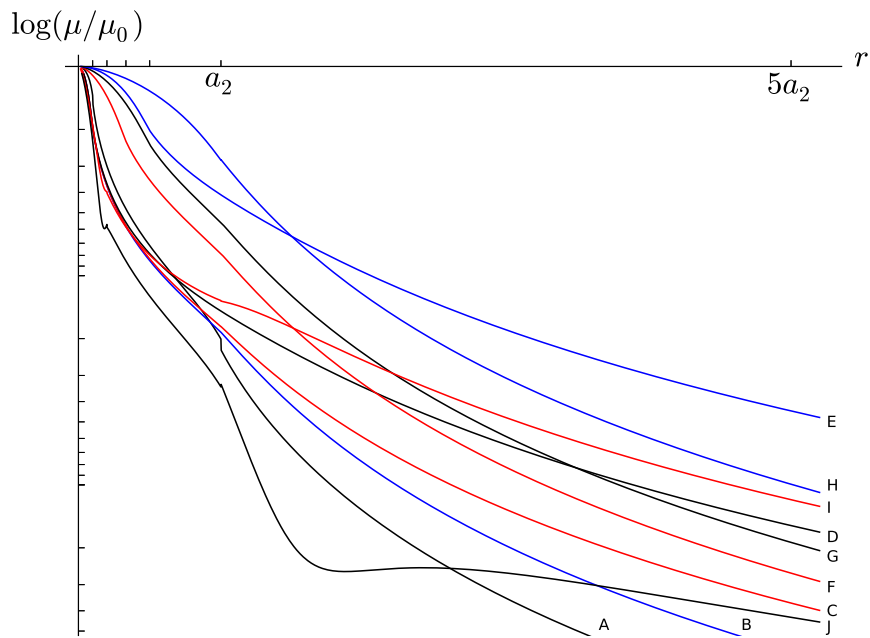}
\caption{The log scaled normalised densities $\mu(r)/\mu(0)$, or ``brightness profiles'' of models A-J in Table \ref{tab:various_disc_model_paramaters}. The smooth, pseudo-exponential behaviour for $r \sim a_2$ can be seen in most profiles, and all profiles eventually become concave upward as $r \to \infty$. The profiles of the extreme models I and J exhibit sharp changes in density, corresponding to regions of sharp change in velocity.}
  \label{fig:P_MODELS_DENS}
\end{figure*}
\clearpage
}
\subsection{Pseudo-exponential behaviour}\label{sec:pseudo-expon-behav}
Finally note that in the intermediate region $a_1 < r < a_2$, and for a period beyond $r=a_2$, the plot of $\log(\mu(r))$ is very close to being linear for the smoother models. Recalling the graph of $\log(e^{-r})$, it can be said that $\mu(r)$ has close to exponential or \emph{pseudo-exponential} behaviour in this region. This results from density having leading order behaviour 
\begin{equation}
 \label{eq:mudisc_flat_approx}  2 \pi G \mudisc(r) \cong \frac{A}{r}-C
\end{equation}
for $a_1 < r < a_2$, which can be derived from (\ref{eq:rho_LR_text}) and (\ref{eq:rho_RL_text}).

Seiden \emph{et al} \cite{Seiden1984} have previously argued that disc brightness profiles are of the form \eqref{eq:mudisc_flat_approx}. Depending on the ratio $a_1/a_2$, the value of $a_2$ itself, and the smoothness of the model, this pseudo-exponential behaviour can continue across several scale lengths. This behaviour eventually breaks down as $r \to \infty$, with the profile becoming $O(1/r^3)$. Such broken-exponential behaviour is suggestive of the phenomenon of truncation in spiral galaxies, first studied by der Kruit\cite{1979A&AS...38...15V,2001ASPC..230..119V}, and observed to be common in spiral galaxies(see Pohlen\cite{Pohlen2006}).

Assuming a constant mass to light ratio, observations of galaxies with profiles of the form \eqref{eq:mudisc_flat_approx} should expect to see pseudo-exponential behaviour so long as the rotation curve remains flat. Since curves remain flat out to most observed distances, indefinitely exponential behaviour is sometimes assumed. However, the models in this paper -- along with observations of truncation -- suggest that (pseudo-)exponential behaviour is tied to the phenomenon of flat rotation curves. And by the results of section \ref{sec:asympt-restr-veloc} below, such curves must eventually decay for real galaxies.

\section{Asymptotic restrictions on velocity}\label{sec:asympt-restr-veloc}

The models above show how quickly and easily disc models can be constructed using this framework. Another significant advantage of the method is that the resulting density profiles are analytic, and so their asymptotic behaviour as $r \to \infty$ can be investigated. Indeed, this behaviour places restrictions on $\vsqr(r)$ if the resulting disc is to have finite total mass so that the model is physically reasonable.

For a given radial density $\mu(r)$, the corresponding disc mass profile $M(r)$ is defined by
\begin{align}
\label{eq:Mdisc_2D_def}  \Mdisc(r) &=  2 \pi \int_0^{\, r} r \mu(r) dr,
%\label{eq:M2dgal_integral} \Mtot &= 2 \pi \int_0^\infty  r \mudisc(r) dr
\end{align}
and the corresponding \emph{total mass} is given by $\Mtot=\lim_{r \to \infty} \Mdisc(r)= 2 \pi \int_0^{\infty} r \mu(r) dr$. For $\Mtot$ to be finite, this improper integral must be convergent; Hence $\mudisc(r)$ must decrease strictly faster than $O(1/r^2)$ as $r \to \infty$. Assuming that $\mudisc$ has no singularities elsewhere in $r$, this asymptotic condition is sufficient to ensure that the galaxy has finite mass.

With this in mind, equation (\ref{eq:4}) can also be used to derive results relating the asymptotic behaviour of the velocity profile $\vsqr(r)$ to that of $\mu(r)$ as $r \to \infty$. These theorems are derived in full in Appendix \ref{sec:asympt-behav-dens}, but for the sake of brevity the results are summarised here in the following theorem.

\begin{thm}
\label{thm:cyl_mass_asym_thms_all}
 For infinitely thin, axisymmetric galactic discs, with velocity profiles(rotation curves) that are asymptotically either: \textbf{I)} Flat \  ($\vsqr(r) \sim C$), \textbf{II)} Keplerian \ ($\vsqr(r) \sim C/r$), or \textbf{III)} rapidly decreasing \ ($\vsqr(r) \sim C/r^n, n>1$), as $r \to \infty$; it follows from   \eqref{eq:1}/\eqref{eq:4} that the corresponding density profiles $\mu(r)$ of the discs have the respective asymptotic behaviours \textbf{I)} $\mu(r) \sim C/r$ , \textbf{II)} $\mu(r) \sim C/r^3$, \textbf{III)} $\mu(r) < 0$, as $r \to \infty$.

Hence from \eqref{eq:Mdisc_2D_def}, assuming that $\mu(r)$ has no other singularities, these conditions mean respectively that \textbf{I)} overall disc mass is infinite ($M(r)$ diverges as $r \to \infty$), \textbf{II)} overall disc mass is finite, \textbf{III)} disc mass in the outer regions is negative.
\end{thm}
In fact, the results of this theorem are exactly equivalent to those for spherical distributions of matter with the same $\vsqr(r)$. Collectively, results \textbf{I)}, \textbf{II)}, \textbf{III)} show that Keplerian $O(1/r)$ velocity profile behaviour at infinity is a \emph{necessary} condition for a physically reasonable disc, that is, one with positive density and finite total mass.

Moreover, there is an additional theorem which places restrictions on the velocity profiles of such discs.
\begin{thm}
\label{thm:mudisc_discont_vel_text}
If the velocity profile $\vsqr$ has a single discontinuity at the point $r=a$, with magnitude
\[
\Delta \vsqr_a = \vsqr(a+0)-\vsqr(a-0)\ne 0
\]
then the resulting radial density $\mudisc(r)$ has leading order asymptotic behaviour of at least
\begin{equation}
 2 \pi G \mudisc(r) \sim -\frac{\Delta \vsqr_a}{r}+O(1/r^3)  
\end{equation}
as $r \to \infty$. Hence by \eqref{eq:Mdisc_2D_def} the total mass of the disc diverges to $\pm \infty$ even if $\vsqr(r)$ subsequently becomes Keplerian at infinity.
\end{thm}
A full proof of this theorem is given in Appendix \ref{sec:proof-theor-refthm:m}. This result can be extended to $\vsqr$ with multiple discontinuities, and with such it is possible for the disc to have overall \emph{negative} mass. This result can be seen as a manifestation of the sensitivity of $\mudisc$ to even small changes in $\vsqr$, and to the ill-posedness of the inverse problem. This behaviour has no corresponding analogue in the spherical galaxy case.

%%%%%%
Theorem \ref{thm:mudisc_discont_vel_text} presents great difficulties if density or mass profiles are to be inferred, e.g. numerically, from velocity profiles measured as discrete data sets. Unless the methods used mollify the discontinuous nature of the measured data, explicitly or implicitly, the resulting mass profiles will diverge as $r \to \infty$. Such difficulties can be avoided by approximating $\vsqr(r)$ using continuous piecewise polynomials. But the use of truncated series for $\vsqr(r)$ requires some care in order to avoid introducing any discontinuities when truncating.
%%%%%

Combined, theorems \ref{thm:cyl_mass_asym_thms_all} and \ref{thm:mudisc_discont_vel_text} require $\vsqr(r)$ to be both continuous and asymptotically Keplerian in order for the resulting disc to be physically reasonable. Therefore, taking the assumption of physically reasonable disc galaxies as a guiding principal, discontinuous velocity profiles, or ones which remain flat indefinitely(e.g. Mestel's Disc) will be rejected in this paper.

\section{Disc mass profiles}\label{sec:disc-mass-profiles-1}

To examine disc mass profiles in more detail, it is useful to introduce elementary mass profiles
\begin{align}
\notag GM_L(r;n) &= \int_0^r r\, 2 \pi G \mu_L(r;n) dr\\
\label{eq:elemen_scaled_mass_profiles}    \text{and} \quad  GM_R(r;n) &= \int_0^r r\, 2 \pi G \mu_R(r;n) dr
\end{align}
based on the corresponding elementary density profiles (\ref{eq:rho_L_def})/(\ref{eq:rho_L_def}). With these, applying \eqref{eq:mudisc_constants_sum} to \eqref{eq:Mdisc_2D_def} allows the total mass profile to be written as the sum
\begin{align}
\label{eq:Mdisc_poly_sum}G \Mdisc(r) = \sum_n \Lcal_n G M_L(r;n) + \Rcal_n G M_R(r;n)
\end{align}
Piecewise series for the elementary profiles $GM_L$ and $GM_R$ are given by the following lemmas, with the constants $W_n$ and $B_n$ as in section \ref{sec:disc-model-framework}. Complete proofs of these lemmas are given in Appendix \ref{sec:disc-mass-profiles}.
%\begin{widetext}
\begin{lem} \label{lem:GML_text}
For $n>0$, with $\vsqr_L(r)=\vsqflat (r/a_1)^n$ giving $\mu_L(r;n)$ as in lemma \ref{lem:muL_series}, the profile $GM_L(r;n)$ is given by
\begin{equation}
  \label{eq:GM_L_text}
  G M_L(r;n)=\begin{cases}
    G M_{LL}(r;n)&, r<a_1\\
    G M_{LR}(r;n) &, r>a_1
    \end{cases}
\end{equation}
where $G M_{LL}$ and $G M_{LR}$ are the left and right parts of $GM_L$, which are given by the series
%\begin{widetext}
\begin{align}
\notag G M_{LL}(r;n) =& \vsqflat  n a_1 \Bigg[\sum_{\begin{subarray}{c}
m=0\\
2m\ne n-1\end{subarray}}^{\infty}\frac{-G_{2m}}{(2m-n+1)(2m+2)}\left(\frac{r}{a_{1}}\right)^{2m+2} \\
\label{eq:GMLL_text}& \quad +\left(\frac{r}{a_{1}}\right)^{n+1}\bigg[\frac{W_{n-1}}{n+1}- \frac{B_{n-1}G_{n-1}}{n+1}\bigg(\ln\left(\frac{r}{a_{1}}\right)-\frac{1}{n+1}\bigg)\bigg]\Bigg]\\
\label{eq:GMLR_text} G M_{LR}(r;n)=&  - \vsqflat n a_1 \sum_{m=0}^{\infty}\frac{G_{2m}}{(2m+n)\left(2m-1\right)}\left(\frac{a_{1}}{r}\right)^{2m-1}
\end{align}
%\end{widetext}
\end{lem}
\begin{lem} \label{lem:GMR_text}
For $n>0$, with $\vsqr_R(r)=\vsqflat (a_2/r)^n$ giving $\mu_R(r;n)$ as in lemma \ref{lem:muR_series}, the profile $GM_R(r;n)$ is given by
\begin{equation}
  \label{eq:GM_R_text}
  G M_R(r;n)=\begin{cases}
    G M_{RL}(r;n)&, r<a_2\\
    G M_{RR}(r;n) + \vsqflat  a_2\, \delta_{1n} &, r>a_2
    \end{cases}
\end{equation}
where $\delta_{1n}$ is the Kronecker delta function, equal to one when $n=1$ and zero otherwise; and where the elementary mass profiles $G M_{RL}$ and $G M_{RR}$ are given by the series
%\begin{widetext}
\begin{align}
\label{eq:GMRL_text} G& M_{RL}(r;n)=\vsqflat na_{2}  \sum_{m=0}^{\infty}\frac{-G_{2m}}{(2m+n+1)(2m+2)}\left(\frac{r}{a_{2}}\right)^{2m+2}\\
\label{eq:GMRR_text} G& M_{RR}(r;n)= \vsqflat n a_2  \times  \begin{cases}
\displaystyle   \sum_{m=0}^{\infty}\dfrac{-G_{2m}}{\left(2m-1\right)^2}\left(\frac{a_{2}}{r}\right)^{2m-1}\hfill , n=1\\
\\
\displaystyle   \sum_{\begin{subarray}{c}
m=0\\
2m\ne n\end{subarray}}^{\infty}\dfrac{-G_{2m}}{(2m-n)\left(2m-1\right)}\left(\frac{a_{2}}{r}\right)^{2m-1} \\
\displaystyle  + \left(\frac{a_{2}}{r}\right)^{n-1}\Bigg[\frac{W_{n}}{n-1}+\frac{B_{n}G_{n}}{n-1}\bigg[\frac{1}{n-1}-\ln\bigg(\frac{a_{2}}{r}\bigg)\bigg]\Bigg]\quad  \hfill , n \ne 1
\end{cases}
\end{align}
%\end{widetext}
\end{lem}
Again, though lengthy, the definitions amount to power series whose behaviour can be analysed as $r \to \infty $. It should be noted that although both $M_{LR}$ and $M_{RR}$ have divergent $O(r)$ terms as $r \to \infty$, these terms will cancel in \eqref{eq:Mdisc_poly_sum} as long as $\vsqr(r)$ is continuous. It is also worth noting the change in behaviour of $M_R$ when $n=1$.

\subsection{Total disc mass}\label{sec:total-disc-mass}
Using \eqref{eq:GM_L_text} and \eqref{eq:GM_R_text} in equation \eqref{eq:Mdisc_poly_sum} for $r > a_2$, and noting the exceptional case of $n=1$ in $M_R$, after the cancellation of $O(r)$ terms\footnote{These cancel as $\sum_n \Lcal_n = \sum_n \Rcal_n = 1$.} the leading order behaviour of $M(r)$ is given by
\begin{align*}
G \Mdisc(r) &= \sum_n \Lcal_n G M_{LR}(r;n) + \Rcal_n \left( G M_{RR}(r;n) +\vsqflat  a_2 \delta_{1n} \right)\\
&= \Rcal_1 \vsqflat  a_2 + O \left( \frac{1}{r} \right)+\cdots 
\end{align*}
Hence taking the limit as $r \to \infty $ gives the total mass $\Mtot$ as
\begin{equation}
\label{eq:GMdisctot_kep_coeff_result}
  G \Mdisctot = \vsqflat \Rcal_1 a_2
\end{equation}
With hindsight, this is not entirely unexpected. Note from equation \eqref{eq:vp_piecewise_polynomial} that $\vsqr_0 \Rcal_1 a_2$ is coefficient of asymptotic Keplerian term in $\vsqr(r)$ as $r \to \infty $. Hence, the velocity profile of a thin disc approaches that of a localised object with the same mass and center, namely
\begin{equation}
  \vsqr(r) \sim  \divfrac{G\Mdisctot}{r}
\end{equation}
Hence total disc mass is determined only by the Keplerian behaviour of $\vsqr$ as $r \to \infty$. Physically reasonable thin discs eventually
behave as localised objects.

\subsection{Asymptotic behaviour of mass}\label{sec:asympt-behav-mass}

It is illuminating to study the rate at which disc mass profiles converge to their total. Using \eqref{eq:GMLR_text} and \eqref{eq:GMRR_text}, the leading and next to leading order behaviours of $M_{LR}$  and $M_{RR}$ are given by
\begin{equation}
  \label{eq:GM_LR_leading_orders}
  GM_{LR}(r;n) = \left( \vsqr_0 r  \right) - \frac{\vsqr_0 n a_1 }{4(n+2)} \left( \frac{a_1}{r} \right)+\cdots
\end{equation}
\begin{equation}
   \label{eq:GM_RR_leading_orders}
  GM_{RR}(r;n) =
  \begin{cases}
    \displaystyle \left(-\vsqr_0 r  \right) + \frac{\vsqr_0 n a_2 }{4(n-2)} \left( \frac{a_2}{r} \right)+\cdots &, n \ne 2\\
\\
    \displaystyle \left(-\vsqr_0 r  \right) + \frac{\vsqr_0 a_2}{2} \left( \frac{a_2}{r} \right) \ln \left( \frac{r}{a_2} \right) +\vsqr_0 a_2 \ln(2) \left( \frac{a_2}{r} \right)+\cdots &,n = 2\\
  \end{cases}  
\end{equation}
In each case, the ellipses denote more rapidly decreasing terms as $r \to \infty $. For the exceptional case of $n=2$ in $M_{RR}$, the value of $W_2$ was obtained from table \ref{tab:w_n_table_text}.

Applying these in \eqref{eq:Mdisc_poly_sum} for $r>a_2$, recalling that $O(r)$ terms will cancel for continuous, Keplerian $\vsqr(r)$, it follows that the leading order behaviour of the complete mass profile is given by
\begin{align}
  \notag G\Mdisc(r) =& G \Mdisctot + \Rcal_2 \frac{\vsqr_0 a_2}{2} \left( \frac{a_2}{r}\right) \ln \left( \frac{r}{a_2} \right) - \frac{\vsqr_0 a_1}{4}\sum_{n}^{}\Lcal_n \left( \frac{n}{n+2} \right) \left( \frac{a_1}{r} \right) \\
    \label{eq:GMdisc_asymtotic_behaviour}&\qquad +\frac{\vsqr_0 a_2}{4} \left[ \sum_{n \ne 2}^{}\Rcal_n \left( \frac{n}{n-2} \right)+4\Rcal_2 \ln(2)  \right] \left( \frac{a_2}{r} \right) +\cdots , \quad \text{as r $\to \infty$.} 
\end{align}
Note that the term $(a_2/r) \ln(r/a_2)$ decreases to zero as $r \to \infty $, but does so more slowly than $(a_2/r)$. Now, since $(a_1/r)$ and $(a_2/r)$ have the same essential $O(1/r)$ rate of decrease, this asymptotic behaviour can be more briefly stated as
\begin{equation}
  \label{eq:asym_behaviour_mass}
G \Mdisc(r) = G \Mdisctot +\Rcal_2\, O\brkt{\frac{\ln(r)}{r}} + O \left( \frac{1}{r} \right)+\cdots
\end{equation}
Thus for physically reasonable galaxies, as $r \to \infty $ their mass profile converges to $G\Mtot$ as $O(1/r)$ if $\Rcal_2=0$, and as $O(\ln(r)/r)$ for the exceptional case when $\Rcal_2 \ne 0$. It is important to note that these rates of convergence can be quite slow in practice. To see this, consider the abstract ``mass fraction'' profile $M(r)/\Mtot=1-a_2/r$, which requires $r$ to reach $100 a_2$ in order for mass to pass $99\%$ of the total -- the function $1/r$ approaches zero, but is in no hurry to do so.

%To see this, consider an abstract ``mass fraction'' profile $M(r)/\Mtot=1-a_2/r$, which converges to $100 \%$. At $r=2 a_2$, $m=50\%$ and so half the remaining mass lies beyond $r=2 a_2$. At $r=10 a_2$, $m = 90\%$. At $r=100 a_2$, $ m = 99\%$. At $r=1000 a_2$, $m(r) = 99.9\%$ -- which is within three digits of accuracy of the total mass, but this required traversing an enormous distance beyond $a_2$. This behaviour ultimately results from the slow convergence of $1/r$; while this function approaches zero, it is in no hurry to do so.
Similar behaviour will hold for $G \Mdisc(r)$. Consider the $\Rcal_2=0$ case, and assume that $a_1 \ll a_2$, so that dividing \eqref{eq:asym_behaviour_mass} by $G \Mdisctot$ then gives the resulting mass fraction profile as
\begin{equation}
  \label{eq:asym_behaviour_mass_percentage}
\divfrac{\Mdisc(r)}{\Mdisctot} \cong  1 - C\brkt{\divfrac{a_2}{r}}+O\brkt{\brkt{\divfrac{a_2}{r}}^3}, \quad r \gg a_2
\end{equation}
for some constant $C$. This function very slowly converges to $100 \%$, with distance $r$ likely needing to exceed significant multiples of $a_2$ in order for $\Mdisc(r)$ to reach even $90 \%$ of the total $\Mdisctot$. And if the $\Rcal_2 \ne 0$ case is considered, an $O((a_2/r)\ln(r/a_2))$ results in \eqref{eq:asym_behaviour_mass_percentage}, and the resulting mass convergence will be even slower\footnote{Significantly slower than $a_2/r$ in practice. It is worth noting that $\Rcal_2 \ne 0$ also gives the slowest convergence rate for spherical galaxy models as well.}. Slow $O(1/r)$ mass convergence is a characteristic feature of self-gravitating discs reconstructed from velocity profiles.

\subsection{Example models} \label{sec:example-models-1}

\afterpage{
\begin{figure*}[t]%[htbp]
  \centering
  \includegraphics[width=0.8\textwidth]{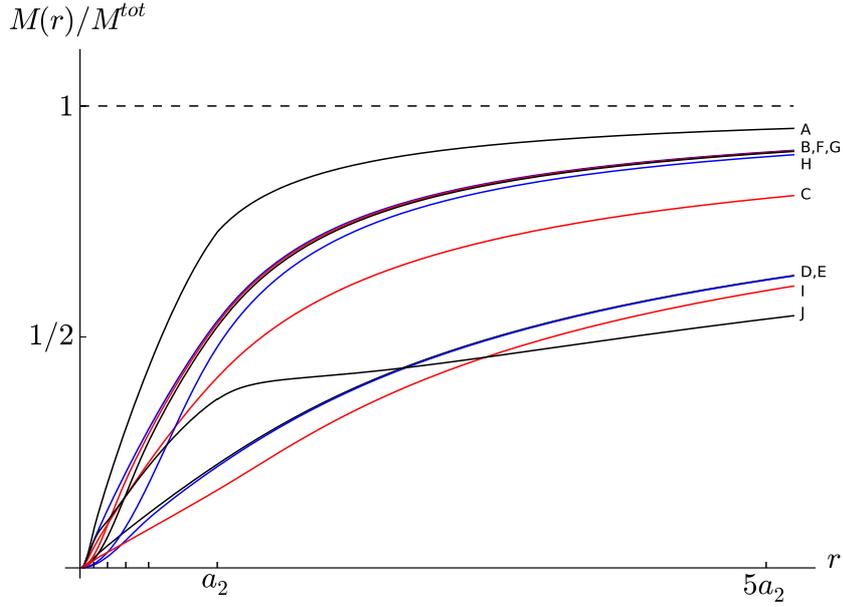}
\caption{Normalised mass fraction profiles $M(r)/\Mtot$ for disc models A-J in Table \ref{tab:various_disc_model_paramaters}. The models display linear behaviour for $r<a_2$, but much slower rates of convergence past this point. Moreover, not only can the mass profile at $r=a_2$ significantly underestimate the total mass, it is in fact possible for the \emph{majority} of the disc's mass to lie out beyond $a_2$.}% The distance $a_2$ is arbitrary, but is equal across all models here for the purposes of comparison.} 
  \label{fig:P_MODELS_MASS}
\end{figure*}

\begin{table}[b]
  \begin{center}
    \renewcommand{\arraystretch}{1.6}
\setlength{\tabcolsep}{0.4em}
    % \begin{tabular}{|c|c|r|r|r|r||c||c|}
    %     \hline
    %     \multirow{2}{*}{Model} & \multicolumn{5}{|c||}{Mass fraction distances} &  \multirow{2}{*}{$\tfrac{\Mdisc(a_2)}{\Mdisctot}$} &  \multirow{2}{*}{$\tfrac{\Mdisctot}{\Mdisctot_A}$} \\
    %      & $\tfrac{\Mdisc(r)}{\Mdisctot}$ & 50\% & 90\% & 95\% & 99\% & &   \\
    %     \hline
    %     A & $r/a_2$ & 0.6& 2.5& 5.1& 25.2& 75.5\% & 1 \\
    %     \hline
    %     B & $r/a_2$ & 0.9& 5.0& 10.0& 50.1& 53.4\% & 3/2 \\
    %     \hline
    %     C & $r/a_2$ & 1.3& 12.2& 28.7& 191.0& 41.2\% & 2 \\
    %     \hline
    %     D & $r/a_2$ & 3.0& 32.8& 78.8& 538.1& 22.4\% & 4 \\
    %     \hline
    %     E & $r/a_2$ & 3.1& 32.9& 78.9& 538.7& 21.9\% & 4 \\
    %     \hline
    %     F & $r/a_2$ & 0.9& 5.0& 10.1& 50.6& 52.9\% & 3/2 \\
    %     \hline
    %     G & $r/a_2$ & 1.0& 5.1& 10.2& 51.3& 52.2\% & 3/2 \\
    %     \hline
    %     H & $r/a_2$ & 1.1& 5.5& 11.0& 55.0& 47.9\%  & 3/2 \\
    %     \hline
    %     I & $r/a_2$ & 3.4& 34.4& 81.9& 552.6& 16.8\%  & 6 \\
    %     \hline
    %     J & $r/a_2$ & 4.0& 59.6& 146.1& 1022.3& 36.5\%  & 2 \\
    %     \hline
    % \end{tabular}
    \begin{tabular}{|c|c|r|r|r|r||c||c|}
        \hline
        \multirow{2}{*}{Model} &  & \multicolumn{4}{|c||}{Mass fractions --  $\tfrac{\Mdisc(r)}{\Mdisctot}$ } &  \multirow{2}{*}{$\tfrac{\Mdisc(a_2)}{\Mdisctot}$} &  \multirow{2}{*}{$\tfrac{\Mdisctot}{\Mdisctot_A}$}\\
         \cline{3-6} &  & 50\% & 90\% & 95\% & 99\% & &   \\
        \hline
        A & \multirow{10}{*}{$r/a_2$} & 0.6& 2.5& 5.1& 25.2& 75.5\% & 1 \\
        \cline{1-1} \cline{3-7}
        B &  & 0.9& 5.0& 10.0& 50.1& 53.4\% & 3/2 \\
        \cline{1-1} \cline{3-7}
        C &  & 1.3& 12.2& 28.7& 191.0& 41.2\% & 2 \\
        \cline{1-1} \cline{3-7}
        D &  & 3.0& 32.8& 78.8& 538.1& 22.4\% & 4 \\
        \cline{1-1} \cline{3-7}
        E &  & 3.1& 32.9& 78.9& 538.7& 21.9\% & 4 \\
        \cline{1-1} \cline{3-7}
        F &  & 0.9& 5.0& 10.1& 50.6& 52.9\% & 3/2 \\
        \cline{1-1} \cline{3-7}
        G &  & 1.0& 5.1& 10.2& 51.3& 52.2\% & 3/2 \\
        \cline{1-1} \cline{3-7}
        H &  & 1.1& 5.5& 11.0& 55.0& 47.9\%  & 3/2 \\
        \cline{1-1} \cline{3-7}
        I &  & 3.4& 34.4& 81.9& 552.6& 16.8\%  & 6 \\
        \cline{1-1} \cline{3-7}
        J &  & 4.0& 59.6& 146.1& 1022.3& 36.5\%  & 2 \\
        \hline
    \end{tabular}
  \end{center}
  \caption{The approximate multiples of $a_2$ at which the disc mass profiles reach given percentages of the total mass for each of the ten models A-J in Table \ref{tab:various_disc_model_paramaters}. For each model the percentage of mass enclosed within $r=a_2$ is also listed, and is typically a minority of the total. The relative mass of each model to that of the disc in model A is also given.}
  \label{tbl:mass_profile_models}
\end{table}
\clearpage
}
Returning to the models of Table \ref{tab:various_disc_model_paramaters}, using \eqref{eq:Mdisc_poly_sum}, \eqref{eq:GM_L_text}, \eqref{eq:GM_R_text}, and \eqref{eq:GMdisctot_kep_coeff_result}, the models' mass fraction profiles $M(r)/\Mtot$ can be computed. The resulting profiles are plotted in Figure \ref{fig:P_MODELS_MASS} for fixed $a_2$.

From the plots, it can first be seen that mass $M(r)$ rises in an approximately linear fashion in the flat velocity region $a_1 < r < a_2$ in most models. This corresponds to the pseudo-exponential behaviour \eqref{eq:mudisc_flat_approx} of $\mu(r)$ in the region -- model H again being the exception. This linear increase is similar to that seen in spherical galaxy models in the same region.

However beyond $r=a_2$ the rate of convergence of $M(r)$ slows considerably, confirming the analysis of section \ref{sec:asympt-behav-mass}. This slow $O(1/r)$ convergence occurs across all models, even in model A, where $\vsqr(r)$ becomes exactly Keplerian for $r>a_2$. In addition, the value of the mass fraction at $r=a_2$ -- the outer edge of flat velocity region -- is generally at around $50\%$ or less, and is as low as $22\%$ in some models, meaning the flat velocity region encompasses a minority of the total mass.

\subsubsection{Outer mass fraction distances}
Table \ref{tbl:mass_profile_models} lists the values of $r/a_2$ corresponding to the $50\%$,$90\%$,$95\%$, and $99\%$ mass fraction values\footnote{The required $r/a_2$ can be found numerically; for example by using Newton's method to find the root of $M(r)/\Mtot -p =0$ for a given percentage $p$.} for each of the ten models A-J. The values are the multiples of $a_2$ required to reach the specified fraction of total mass for that model. Also listed are the flat region edge mass fractions $\Mdisc(a_2)/\Mdisctot$ for each model, and also the total mass of each model relative to the mass of model A.

The table values illustrate the slow convergence of disc mass beyond the flat velocity region -- beyond $a_2$. The first smooth velocity model, B, requires $r\cong 5a_2$ to reach $90 \%$ of total mass; C requires $r \cong 12a_2$ and D requires $r \cong 32 a_2$. The situation worsens at higher mass percentages. To reach the $99 \%$ mark, even the non-smooth model A requires $r>25a_2$, and the slowly decreasing model D requires distance to reach a staggering $r>538\, a_2$. Since the flat velocity length scale $a_2$ may be measured in tens of kiloparsecs($\sim 3 \times 10^{19}$ meters), $~500a_2$ can be of the order of $10^{22}$ meters, which is the scale of intergalactic distances.

Note that model groups D,E and B,F,G,H respectively have the same $\vsqr_R(r)$, but different $\vsqr_L(r)$ and ratios $a_1/a_2$. Comparing their mass fraction distances show that their asymptotic behaviour depends essentially on $\vsqr_R(r)$. Changes in $\vsqr_L(r)$ or the ratio $a_1/a_2$ have only minor effects on the rate of convergence, and no effect on the total mass by (\ref{eq:GMdisctot_kep_coeff_result}). Finally, note that the models with the largest mass fraction distances are those for which $\Rcal_2 \ne 0$, reflecting their slow $O(\ln(r)/r)$ rates of convergence.%It is possible that this rate is physically significant, or indeed physically prohibited.

Before investigating outer mass further, it should be noted that models can agree on the \emph{inner} mass of the galaxy if their inner velocity profiles are similar. Hence it is possible for disc profiles reconstructed from the same incomplete velocity measurements to agree on mass within the inner or measured region, but to disagree completely on the total disc mass(see Kostov \cite{Kostov2007}). The only way to find total mass is to measure beyond $r=a_2$ to determine the asymptotic Keplerian behavior of the velocity profile(see Honma and Sofue \cite{1996PASJ...48L.103H}).

\subsection{Disc hinterlands and the problem of missing mass} \label{sec:relev-probl-miss}\label{sec:disc-hint-probl}

The disc models A-J show that the edge, $r=a_2$, of the flat velocity region in no way signifies the outer limits of a disc galaxy. It can be said that planar disc galaxies possess substantial mass ``hinterlands'': Large regions coplanar with the disc, far from the axis of rotation and of sparse density, but that can contain significant fractions or even the \emph{majority} of the galaxy's total mass (see Figure \vref{fig:gal_hinterlands}, left). The term hinterland has been chosen here to emphasise the two dimensional nature of these mass distributions, in contrast to the three dimensional spherical geometry of standard galaxy halos\footnote{It is notable that spherical models can also possess such slow mass convergence in their halos, but only if $\vsqr(r)$ has an $O(1/r^2)$ asymptotic term as $r \to \infty$. That is if $\Rcal_2 \ne 0$ (See section \eqref{sec:asympt-behav-mass}). }.

This result is in some sense counterintuitive, suggesting that most of the mass in disc galaxies may lie in regions presently considered to be beyond their more obvious bounds. Such a mass distribution can be compared to the power distribution(encircled energy) of Airy patterns in optics. These also possess slow $O(1/r)$ convergence rates as $r \to \infty$(see  \cite{born_and_wolf}), resulting in large fractions of energy lying some distance from the central Airy disc.% However, image patterns with quite low intensity central Airy discs must be considered to compare with some of the disc models shown above.

It should be noted that hinterlands have quite rapidly decreasing densities, $\mudisc(r) \sim O(1/r^3)$, and so it may be difficult to detect or even recognise the presence of matter there particularly if density is too low to support star formation. Additionally, in the present formalism hinterlands do not begin until $\vsqr(r)$ begins to decline. Since most spiral galaxies are measured to have flat rotation curves out to all observed distances(see \cite{Rubin1980}), existing measurements therefore do not extend into potential hinterland regions, and so offer no observational evidence for their existence.

\subsubsection{Hinterlands as artifacts vs real phenomena}\label{sec:hint-as-artif}

This phenomenon of disc hinterlands followed entirely from the $O(1/r)$ converge of mass profiles which were reconstructed from velocity profiles using (\ref{eq:1}). But hinterlands are not a universal feature of all thin disc models. It is possible to simply consider truncated density profiles with $\mu(r)=0$ for large $r$, or else exponentially decreasing profiles, $\mu(r) \sim \mu_0 e^{-r/a}$, such as Freeman's disc\cite{Freeman1970} or indeed Toomre's original\footnote{Toomre\cite{1963ApJ...138..385T} noted that solutions to (\ref{eq:1}) tended to produce slow $O(1/r)$ convergence, but felt that such behaviour was not likely to representative of actual galaxies.} Gaussian mass distribution\cite{1963ApJ...138..385T}. Such models do not posses hinterlands of any kind, and their mass profiles behave in a more localised way -- though their rotation curves are not flat.

The velocity profiles of such discs can be inferred from the forward integral of Toomre\cite{1963ApJ...138..385T} corresponding to \eqref{eq:1}, namely
 \begin{equation}
   \label{eq:toomre_v2_forward}
 \frac{\vsqr(r)}{2 \pi Gr}= \inthinf  \rhat \mu(\rhat) \inthinf k J_0(k\rhat) J_1(kr)  dk d\rhat 
 \end{equation}
Such $\vsqr(r)$ will have Keplerian behaviour at infinity as expected. If these velocity profiles are measured and used in \eqref{eq:1} to reconstruct $\mu(r)$ and $M(r)$, then owing to the inevitable presence of error in both observation and any numerical calculation, the inferred mass profile will contain residual $M(r) \sim O(1/r)$ terms as $r \to \infty$. Appendix \ref{sec:effect-veloc-pert} quantifies this, showing that the change in the rate of asymptotic convergence of mass is given by 
\begin{equation}
  \label{eq:delta_mass_convergence_rate_change}
  \Delta M(r) = \frac{1}{2 \pi G r} \left[ 2 \pi \int_{p_1}^{p_2} \frac{1}{2} \Delta \vsqr(\rhat) \rhat d \rhat \right] + O \left( \divfrac{1}{r^3} \right)
\end{equation}
as $r \to \infty$, where $\Delta \vsqr$ is a perturbation in the velocity profile, and $[p_1,p_2]$ is the finite interval over which $\Delta \vsqr$ is non-zero. And so the reconstructed disc may possess a mass hinterland which the original disc did not.

This demonstrates that is possible for hinterlands to be an entirely artificial residue -- an artifact -- of the inversion process. This behaviour is a consequence of the ill-posedness of the inverse problem in \eqref{eq:1}. Thus whether hinterlands are physically real or not, they are an \emph{inevitable} artifact in the inversion of disc velocity profiles.

\subsubsection{Cutting off Hinterlands}

It is possible to simply introduce a cutoff in inferred density profiles $\mu(r)$ to restrict such hinterlands. Doing so does not significantly impact the resulting inner rotation curve, but does affect the outer curve and overall mass.

\begin{figure*}[b]%[htbp]
  \centering
  \includegraphics[width=0.8\textwidth]{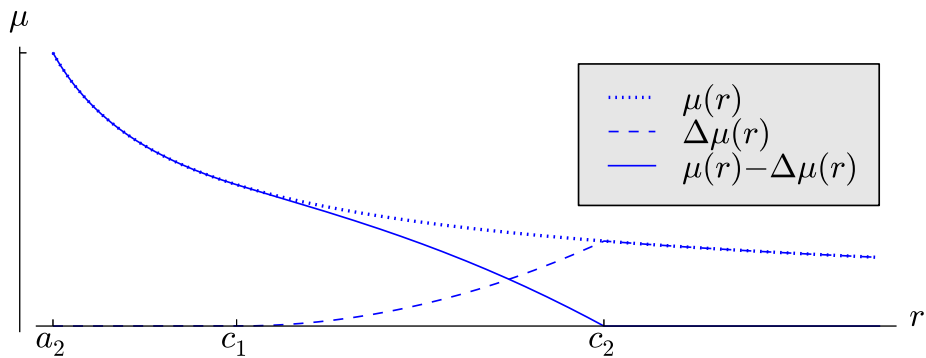}
\caption{A cutoff of an inferred density profile $\mu(r)$. The cutoff function $\Delta \mu$ begins at $r=c_1$, and the resulting density $\mu-\Delta \mu$ is zero past $r=c_2$. Such a cutoff removes hinterlands without seriously affecting flat velocity profiles, but can lead to significant changes in overall mass.}
  \label{fig:density_cutoff_plot}
\end{figure*}
Consider a continuous density cutoff function $\Delta \mu(r)$, non-zero for $r > c_1 > a_2$ and equal to $\mu(r)$ for $r> c_2$ (See Figure \vref{fig:density_cutoff_plot}). Though not proven here, it can be shown that if such a cutoff function is subtracted from $\mu(r)$, then using \eqref{eq:toomre_v2_forward} and methods similar to those used in section \ref{sec:disc-model-framework} allows asymptotic behaviour and bounds for the change $\Delta \vsqr$ in the velocity profile for $r < a_2$ to be derived as
\begin{align}
  \label{eq:8}
  \Delta \vsqr (r) &= 2 \pi G r^2 \left[ \frac{1}{2} \int_{c_1}^\infty \frac{\Delta \mu(\rhat)}{\rhat^2} d \rhat \right] + 2\pi G r O \left( \left(   \divfrac{r}{c_1} \right)^3 \right)\\
  \label{eq:8_2} & \le r^2 \left[ \frac{ 2 \pi G |\Delta \mu|_{\text{max}}}{2c_1} \right] +  r\, O \left( \left(   \divfrac{r}{c_1} \right)^3 \right)
\end{align}
where $|\Delta \mu|_{\text{max}}$ is the maximum value of the density cutoff. In fact, letting $|\Delta \mu|_{\text{max}}= \epsilon \mu(0)$ and $c_1=\alpha a_2$, it can be further shown that
\begin{equation}
  \label{eq:9}
  \Delta \vsqr (r) \le \vsqr_0 \left( \frac{r}{a_2} \right)^2 \frac{\epsilon}{2\alpha} + r O \left( \left(   \divfrac{r}{c_1} \right)^3 \right)
\end{equation}
Experimental results suggest that typically, for $\alpha = 2$, $\epsilon  \sim  10^{-2}$ or less. Thus in its inner regions the velocity profile is affected by at most a quadratic term with a relatively low coefficient, so that the flat velocity profile is relatively unchanged. Hence hinterlands can be cut off while still preserving a relatively constant rotation curve.

However, such an option is not available if the rotation curve has been measured out to a point where it's Keplerian decline is known. By \eqref{eq:GMdisctot_kep_coeff_result}, the mass of the galaxy is determined completely by the Keplerian rate of $\vsqr(r)$ at infinity. But as seen from the mass fractions in Table \ref{tbl:mass_profile_models}, a density cutoff $c_1/a_2 \sim 2$ can reduce the total mass considerably -- even in model A, such a cutoff can decrease mass by more than $10\%$.

Hence cutoffs are only feasible in cases where the outer rotation curve is unknown, which allows for some latitude when modelling. In cases where the outer rotation curve has been measured, the nature of its transition from flat to Keplerian behaviour fixes the extent and the significance of the galaxy's hinterlands.

\subsubsection{Retaining Hinterlands}\label{sec:reta-hint}

Ultimately the actual existence of such hinterlands can only be determined by empirical investigation of outer rotation curves. It is entirely possible that measured curves will simply resemble those of more localised discs. However, the presence of an extended hinterland could offer a new perspective on the problem of missing matter in galactic clusters, and so it may be worth exploring whether these assumed artifacts could be physically real.

It seems difficult to avoid some kind of hinterland when using disc models. Model A suggests that a rapid transition to Keplerian behaviour will lead to a ``minimal'' hinterland with perhaps only $25\%$ of total mass. And as shown in section \ref{sec:asympt-behav-mass}, the presence of an $O(1/r^2)$ asymptotic term in $\vsqr(r)$ can lead to this mass being spread over an extremely wide region; this is the $\Rcal_2 \ne 0$ case seen in models C,D,E,I, and J above. %It is also possible to consider discs which possess temporary $O(1/r)$ mass convergence before later becoming more localised or exponential.

Such extended hinterlands could play a role similar to extended spherical halos when modelling galaxies and galactic clusters. Like halos, sparse extended hinterlands can explain the discrepancies between the measured masses of clusters and those expected from their visible matter. In addition, hinterlands do this with distributions of matter that lie in the same plane as the observed matter of the galaxy. Moreover, the non-spherical nature of hinterlands may affect the expected dynamics of clusters, as discs cannot be approximated as point masses, leading to greater or smaller than expected forced between galaxies.

Given such possibilities, it may be worth retaining extended hinterlands when modelling instead of going to the trouble of cutting them off as an assumed artifact.

\section{Numerical simulation}\label{sec:galaxies-gas-nebulae}

\begin{figure*}[t]
\begin{center}
   \includegraphics[width=0.43\textwidth]{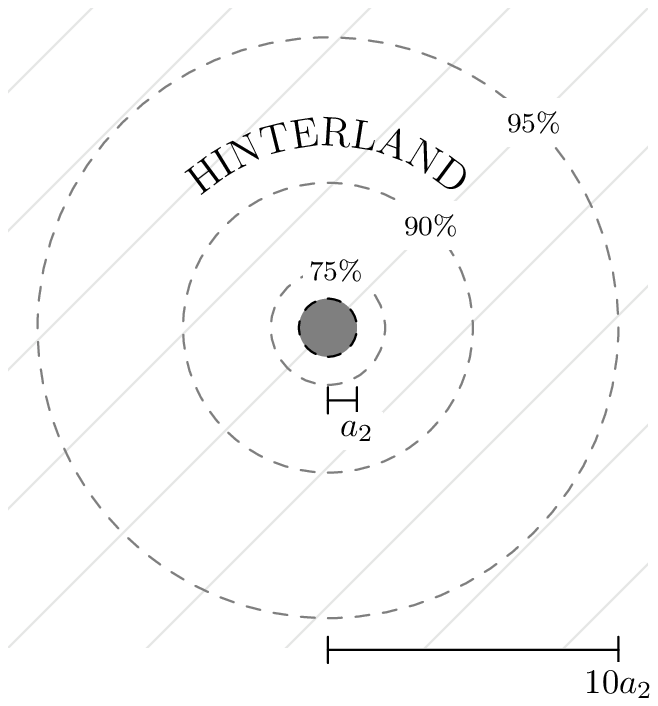}
   \includegraphics[width=0.50\textwidth]{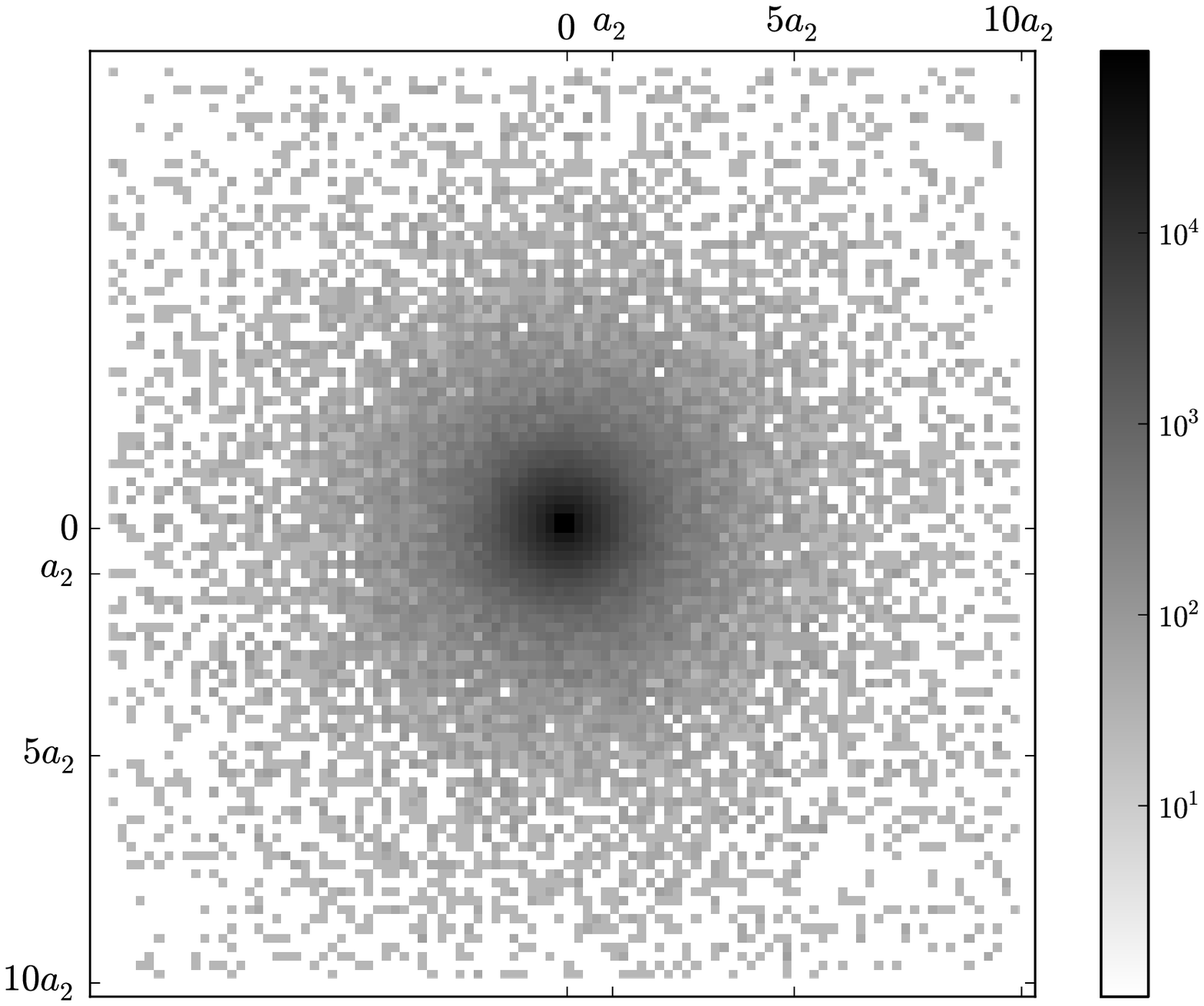}
\end{center}
\caption{
\indent \textbf{Left:} The mass contours of disc model B. Each circular contour gives the percentage of the total galactic mass lying inside it. The central portion of the galaxy (grey) contains only $\sim 53\%$ of the total mass, with almost half the galaxy's mass lying in a "hinterland" (hatched) beyond the flat velocity region, $r>a_2$. Increasingly vast areas, ever farther from the center are required to encompass higher mass percentages. \\
\indent \textbf{Right:} Plot of the simulated galaxy of $10^6$ masses based on model B. Density per unit area is shown on a logarithmic scale. The simulation confirms that a large fraction of disc mass lies beyond $a_2$. The area of the plot contains $\sim 95\%$ of the total masses generated.\\
}
\label{fig:gal_plot_log_10}
\label{fig:gal_hinterlands}
\label{tab:model_B_hinterland_plots}
\end{figure*}
%It is shown that if a galaxy generated to match a density profile $\mudisc$ will naturally reproduce the original velocity profile $\vsqr$ from which $\mu$ was derived..

To confirm the analysis of sections \ref{sec:disc-model-framework} and \ref{sec:disc-mass-profiles-1}, a simplified numerical check is now presented. A static simulated galaxy of $N=10^6$ homogeneous nebulae is generated to conform to the density profile of the disc model $B$ of section \ref{sec:example-models}. The resulting averaged forces in the simulation match those expected to support the corresponding velocity profile.

Using the parameters of model $B$ given in Table \ref{tbl:mass_profile_models}, equations \eqref{eq:vp_piecewise_polynomial} and \eqref{eq:rhodisc_polynomial} give the velocity/density profile pair
\begin{equation}
  \label{eq:model_2_velocity}
 \vsqr(r) = \vsqflat \times \begin{cases}
2(r/a_1)^2-(r/a_1)^4 &, r<a_1 \\
1 &, a_1<r<a_2 \\
\frac{3}{2}(a_2/r)-\frac{1}{2}(a_2/r)^3 &, a_2<r
\end{cases}
\end{equation}
\begin{align}
\notag 2&\pi G \mudisc(r)=\vsqflat \times  \\
\label{eq:model_b_density_piecewise_series}&\begin{cases}
\displaystyle \sum_{m=0}^\infty \frac{4}{a_1}\left[ \frac{-G_{2m}}{2m-1}+\frac{G_{2m}}{2m-3}  \right] \brkt{ \frac{r}{a_1}}^{2m} +\frac{3}{2a_2}\left[ \frac{-G_{2m}}{2m+2}+\frac{G_{2m}}{2m+4} \right] \brkt{ \frac{r}{a_2}}^{2m}\\ \hfill , r<a_1 \\
\displaystyle \sum_{m=0}^\infty \frac{4}{a_1} \left[ \frac{G_{2m}}{2m+2}-\frac{G_{2m}}{2m+4} \right]\brkt{ \frac{a_1}{r}}^{2m+1} +\frac{3}{2a_2}\left[\frac{-G_{2m}}{2m+2}+\frac{G_{2m}}{2m+4}\right] \brkt{ \frac{r}{a_2}}^{2m}\\ \hfill , a_1 < r<a_2 \\
\displaystyle \sum_{m=1}^\infty \frac{4}{a_1} \left[ \frac{G_{2m}}{2m+2}-\frac{G_{2m}}{2m+4} \right]\brkt{ \frac{a_1}{r}}^{2m+1} +\frac{3}{2a_2}\left[ \frac{G_{2m}}{2m-1}-\frac{G_{2m}}{2m-3}\right] \brkt{ \frac{a_2}{r}}^{2m+1}\\ \hfill  , a_2<r
\end{cases}
\end{align}
Using $\mu(r)$ and its corresponding cumulative probability distribution $M(r)/\Mtot$, the centers of the $10^6$ nebula were randomly distributed in a axisymmetric way on a flat 2D plane so as to conform to \eqref{eq:model_b_density_piecewise_series}. A plot of the surface density of the resulting simulated galaxy is shown in Figure \ref{fig:gal_plot_log_10} (right).

As a softening measure -- and to emphasise global attractions over local forces -- each nebula is considered to be a spherically symmetric gas cloud of fixed radius $d_0$, and constant density $\rho_0$, which then has mass $m_0=\rho_0 (\divfrac{4\pi}{3})d_0^3$. The gravitational acceleration $g$ due to such a cloud is given by
\begin{equation}
  \label{eq:nebula_accel_equation}
  g=g(r)=
  \begin{cases}
    \displaystyle \left( \divfrac{G m_0}{d_0^3} \right)r &,\  r<d_0\\
    \displaystyle  \divfrac{G m_0}{r^2} &,\  r>d_0
  \end{cases}
\end{equation}
where $r$ is distance from the centre of the nebula. Note that the total mass of a galaxy of $N$ such particles is $\Mtot=m_o N$, and by \eqref{eq:GMdisctot_kep_coeff_result}, the velocity parameter of such a model must then satisfy $\vsqr_0 = \frac{G m_0 N}{a_2 \Rcal_1}$.
\begin{figure}%[htbp]
  \centering
  \includegraphics[width=0.95\textwidth]{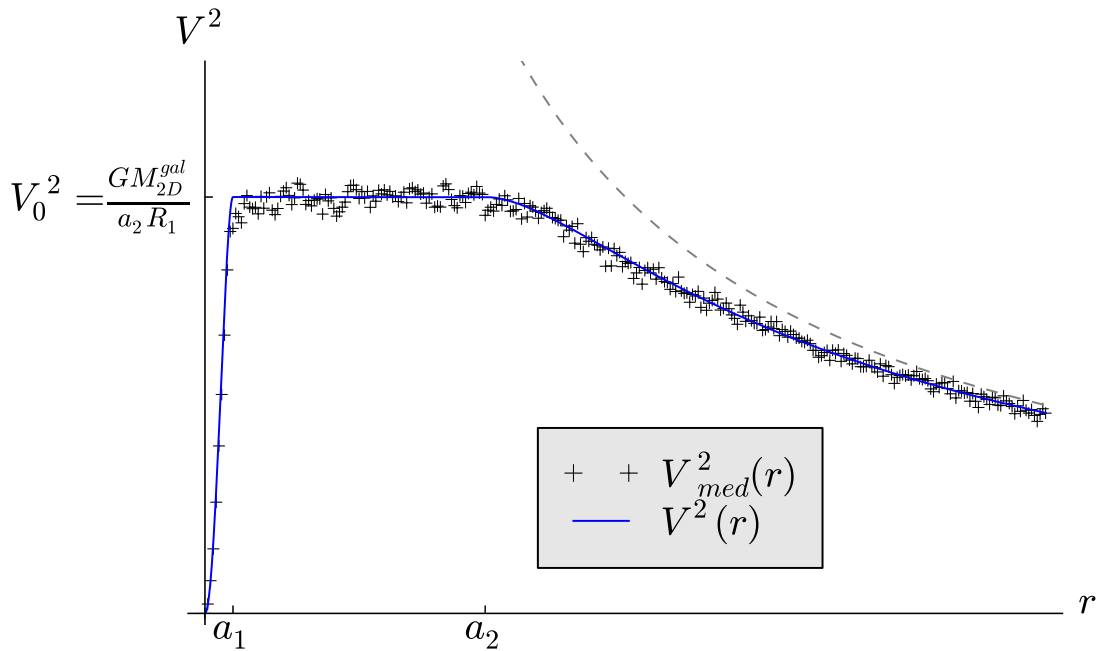}
  \caption{A plot of the numerical velocity profile $V^2_{\text{med}}$ as a function of $r$ for a galaxy of $N=10^6$ masses. The galaxy has been generated to match the radial density \eqref{eq:model_b_density_piecewise_series} of model B in table \ref{tbl:mass_profile_models}, and successfully reproduces the velocity profile of the same. This velocity profile $\vsqr(r)$ given by \eqref{eq:model_2_velocity}, and a Keplerian velocity profile of a point mass $\Mdisctot$ at $r=0$ are also plotted.}
  \label{fig:numerical_velocity_profile_plot}
\end{figure}

Under the approximation of circular orbits about the origin, with centripetal acceleration $a_c$, orbital velocity $V$ is given by $\vsqr = r a_c$. Let $r_i$ be the position of the $i^{th}$ nebula and $a_i$ its total acceleration due to the rest of the galaxy. Then the approximation $V_i$ to its circular orbital velocity is taken to be
\begin{equation}
  \label{eq:num_velocity_model}
  \vsqr_i = \vec{r}_i \cdot \vec{a}_i
\end{equation}
With these individual approximations, consider a circular band of width $\Delta r$ centered at $r$, and the nebulae whose centers lie within this band. The average velocity $\vsqr_{\text{med}}$ at $r$ was taken to be the median\footnote{The median is chosen rather than the mean due to the high variation of individual $a_i$ within each band.} of the approximated velocities $\vsqr_i$ of all nebulae within the band. Though crude, this static approximation gives an estimate of the average central forces experienced in such a mass distribution.

When plotted in Figure \ref{fig:numerical_velocity_profile_plot}, the resulting pairs $ \left( r,\vsqr_{\text{med}} \right)$ closely resemble the expected velocity profile \eqref{eq:model_2_velocity} of the model. The Keplerian velocity profile of a point mass $\Mtot=m_0 N$ at $r=0$ is also plotted. A plot of the corresponding rotation curve $V_{\text{med}}$ is shown in \ref{fig:P_VEL_MODEL_PLOT}(right). This result confirms that the pair \eqref{eq:vp_piecewise_polynomial} and \eqref{eq:rhodisc_polynomial} is a solution to the inverse problem (\ref{eq:1}).

\section{Summary} \label{sec:summary}

This paper has introduced a highly flexible, analytic, thin disc model for self-gravitating axisymmetric galaxies that is capable of describing discs with very general rotation curves. The model was based on the inverse Bessel function method of Toomre, which reconstructs disc surface density from its observed rotation curve. The model also demonstrates that discs alone can support perfectly flat rotation curves of arbitrary finite length, without needed any other galaxy components or modified gravity.

Once a piecewise polynomial expression for the velocity profile $\vsqr$ is given, analytic power series for the disc density and mass profiles, $\mu$ and $M$, are obtained from the polynomial coefficients and known elementary functions. Though these elementary functions are somewhat lengthy, they can be easily evaluated on modern computers and provide much of the flexibility of direct numerical methods at a fraction of the computational cost. Several expository models were constructed to demonstrate the utility of this framework. These examples also revealed the connection between flat rotation curves in discs and the (pseudo-)exponential behaviour of their disc density profiles.

The analytical nature of the resulting series also provides an opportunity to investigate the mathematical relationships between velocity and density profiles of such galaxies. Such investigations lead to theorems which related the asymptotic behaviours of velocity $\vsqr$ to density $\mu$ at infinity. In particular it was shown that requiring the galaxy to be physically reasonable, with finite total mass and everywhere positive density, places restrictions on the velocity profile, requiring it to be continuous and also asymptotically Keplerian at infinity. It is moreover possible to relate the asymptotic behaviours of $\vsqr$ and $\mu$ at the origin, though this was not done in this paper.

The analytical nature of the models also allowed the resulting reconstructed mass profiles to be studied. The mass profiles showed overall linear increase in mass in the flat velocity region, not dissimilar to what would be seen in spherical models. However, further analysis revealed ubiquitous slow $O(1/r)$ convergence of mass in the outer regions of galaxies reconstructed using Toomre's method. This lead to the phenomenon of large, widely distributed, planar mass hinterlands beyond the flat velocity region, which could contain large fractions of even the majority of the galaxy's total mass. The possibility of such hinterlands being just an artifact of inversion was discussed, and the feasibility of simply cutting off the outer regions was examined. However, given the ubiquity of such distributions in reconstructed discs, and their potential to explain problems of missing mass in galactic clusters, there are arguments for retaining hinterlands rather than going to the trouble of removing them.

These results demonstrate the power and flexibility of thin-disc models in describing spiral galaxies with very general rotation curves. The results also offer and alternative perspective on the problem of missing or dark matter. On the scale of individual galaxies, disc models can explain flat rotation curves under Newtonian gravity, and on the larger scale of galaxy clusters, the planar hinterlands of discs can play the role of spherical halos in accounting for the missing matter in such structures. It is hoped that the model framework presented here will be useful in exploring these and other problems in the study of spiral galaxies.

\subsection{Limitations and Future Work}

There are naturally several limitations to the simplified disc models presented in this paper. These limitations suggest future avenues of research.

To begin, only infinitely thin discs have been considered. However it relatively straightforward to modify Toomre's method to consider axisymmetric discs with realistic thicknesses. Again in cylindrical coordinates, let the three dimensional density of the disc be given by $\rho(z)=\mu(r) f(z)$, where $\mu$ is surface density as before, and $f$ is a ``thickness'' function dependant on height $z$ above the galactic plane. In this case surface density in the plane $z=0$ is related to disc velocity $\vsqr$ by 
\begin{equation}
  \label{eq:1_thickness_mod}
  2 \pi G \mu(r) = \int_{0}^{\infty } \frac{d \vsqr(\rhat)}{d \rhat}\int_{0}^{\infty } \left( \tilde{f}(k)\right)^{-1} J_0 \left( k \rhat \right) J_0 \left( k r \right)  dk d\rhat
\end{equation}
where $\tilde{f}$ is equal to twice the Laplace transform of $f$. When $f(z)=\delta(z)$, this reduces to equation \eqref{eq:1} for an infinitely thin disc. Research on realistic thickness profiles is needed, though exponential forms like $f(z)=e^{-\alpha|z|}$ give results which are very amenable to analysis. In addition to thick discs, it may be possible to modify Toomre's method to consider non axisymmetric discs. Such modifications produce kernels similar to (\ref{eq:2}), but more research is needed to prove that these give convergent expressions for density.

As already noted, questions of stability have not been addressed at all in this paper, both for the sake of brevity and because the model is intended as a preliminary investigation of flexible disc galaxy frameworks. The unnaturally thin symmetric discs described in this paper are unlikely to be stable\cite{Toomre1964}, so studies of thick, non-symmetric discs may also be useful in answering questions of stability.

Finally, the velocity profiles which have been studied, of the form (\ref{eq:vp_piecewise}), are rather restrictive. While the purpose of the constant $\vsqr_M$ section was to show that discs alone could support flat rotation curves, in practice slowly rising or falling sections are also seen. Related to this is the problem of fitting $\vsqr_L$ and $\vsqr_R$; Very high order polynomials could be needed in practice. A method which approximates $\vsqr$ using more piecewise intervals and lower degree polynomial splines may be of more practical benefit, though correspondingly more elementary density functions would be required. Future work is also needed on deriving conditions to ensure that densities inferred using Toomre's method are strictly positive.

This work was carried out at the University of Limerick. The research has also made extensive use of NASA's Astrophysics Data System Bibliographic Services, and the SAGE computer algebra system.  

% This paper has shown that it is possible for thin discs of matter alone to support flat rotation curves of arbitrary length under Newtonian gravity. A model framework was presented which gave analytical expressions for disc surface corresponding to piecewise polynomial disc velocities. These models are capable of being quickly and easily fit to observed data.

% This framework was also used to investigate the asymptotic properties of disc mass profiles, leading to necessary restrictions on the velocity profiles of physically reasonable discs. The slow convergence of disc mass was also investigated, showing that discs possess vast mass ``hinterlands'' of increasingly sparse density, but which can contain the majority of their total mass.

% The resulting distributions of matter offer an alternative perspective on the problem of missing or dark matter. On the scale of individual galaxies, disc models can explain flat rotation curves under Newtonian gravity, and on the larger scale of galaxy clusters the planar hinterlands of discs can play the role of spherical halos in explaining the problem of missing mass in these structures. It is hoped that the analytic framework presented can simplify the application of discs models to the investigation of these problems.

%Note that only axisymmetric thin discs have been considered in this paper. Moreover, no analysis of accelerations off the galactic plane $z=0$ has been conducted, nor have problems of rotational stability\cite{Toomre1964} been addressed. 

%\subsection{Acknowledgments}\label{sec:acknowledgments}

\bibliographystyle{plain}
%\bibliographystyle{natbib} 
%\footnotesize{
%\bibliographystyle{mn2e} 
%\bibliographystyle{natbib}
\bibliography{niall-ryan-mathematics} 
%}

\appendix 
\footnotesize

\section{Elementary Density Profiles}\label{sect:app_elem_den_results}

This appendix gives a method for finding the series representations of the elementary density profiles $\mu_L(r;n)$ and $\mu_R(r;n)$, given in Lemmas \ref{lem:muL_series} and \ref{lem:muR_series}. Before this, it is noted that expressions for the constants $W_n$ defined in \eqref{eq:Wn_Def_text_sum} can be found using previous results of Adamchik\cite{VSACatalan2002}, who gave results for series of the form
\[ \sum_{\substack{k=0\vspace{0.1em}  \\k\ne -r }}^\infty \frac{G_{2k}}{k-r}\]
Using these results, it is possible to show that the $W_n$ can also be given by the formula
\begin{equation}
\label{eq:w_n_log_def}
W_{n}=2 G_{n}B_n\left[\ln(2)-\sum_{l=1}^{n} \frac{(-1)^{l+1}}{l} \right] 
\end{equation}
where $B_n=0$ for odd $n$, and $B_n=1$ for even $n$. Thus $W_n=0$ for odd $n$, and in particular $W_1=0$. It is also noted here without proof that $W_n>0$ for even $n$.

Returning to $\mu_L(r;n)$ and $\mu_R(r;n)$, consider their definitions in \eqref{eq:rho_L_def} and \eqref{eq:rho_R_def}. For brevity, define the terms $\rLmin=\text{min}(r,a_1)$ and $\rRmax=\text{max}(r,a_2)$, and split the integrals over $\int_0^{a_1}$ and $\int_{a_2}^{\infty}$ into pairs over $\int_0^{a_<}$, $\int_{a_<}^{a_1}$ and $\int_{a_2}^{a_>}$, $\int_{a_>}^{\infty}$ respectively. Then, noting that the terms $r_<=\text{min}(r,\rhat)$ and $r_>=\text{max}(r,\rhat)$ take on fixed definitions over the resulting non-vanishing intervals, it follows that
\begin{align}
\label{eq:def_muL_r} 2 \pi G \mu_L(r;n) =\sum_{m=0}^{\infty}& \frac{G_{2m}}{r^{2m+1}} \int_0^{\rLmin} \frac{d \vsqr_L(\rhat)}{d \rhat} \rhat^{2m} d \rhat +G_{2m}\, r^{2m} \int_{\rLmin}^{a_1} \frac{d \vsqr_L(\rhat)}{d \rhat} \frac{1}{\rhat^{2m+1}} d \rhat 
\end{align}
\begin{align}
\label{eq:def_muR_r}  2 \pi G \mu_R(r;n) =\sum_{m=0}^{\infty}& \frac{G_{2m}}{r^{2m+1}} \int_{a_2}^{\rRmax} \frac{d \vsqr_R(\rhat)}{d \rhat} \rhat^{2m} d \rhat+G_{2m}\, r^{2m} \int_{\rRmax}^{\infty} \frac{d \vsqr_R(\rhat)}{d \rhat} \frac{1}{\rhat^{2m+1}} d \rhat
\end{align}

The second of these expressions will be used to prove Lemma \ref{lem:muR_series}; the proof of Lemma \ref{lem:muL_series} follows similarly. Now, in Lemma \ref{lem:muR_series} $\vsqr(r)=\vsqflat (a_2/r)^n$ and so $\divfrac{d \vsqr}{d r}=-\vsqflat (n/a_2)(a_2/r)^{n+1}$. And since indefinitely flat rotation curves are not considered, $n>0$. Also recall by \eqref{eq:rho_R_piecewise_text} that $\mu_R(r;n)$ has definition $\mu_{RL}(r;n)$ or $\mu_{RR}(r;n)$ depending on whether $r$ is less than or greater than $a_2$.

Firstly consider the case of $\mu_R=\mu_{RL}$, for which $r<a_2$ and so $a_>=a_2$. Using this in \eqref{eq:def_muR_r} causes the first integral to vanish. Now $n>0$, and so $-n \ne 2m+1$ and thus no logarithmic terms result from the second integral. Hence
\begin{align*}
2 &\pi G \mu_{RL}(r;n)=\sum_{m=0}^{\infty}G_{2m} r^{2m} \int_{a_2}^{\infty} -\vsqflat \frac{n}{a_2} \left( \frac{a_2}{\rhat} \right)^{n+1} \frac{1}{\rhat^{2m+1}} d\rhat\\
=&-\vsqflat \frac{n}{a_{2}}\sum_{m=0}^{\infty}\frac{G_{2m}}{2m+n+1}\left(\frac{r}{a_{2}}\right)^{2m}
\end{align*}
Which is \eqref{eq:rho_RL_text} as required.
%\begin{widetext}

Next consider $\mu_{RR}$, for which $r>a_2$ and so $a_>=r$. In this case, both integrals are present in \eqref{eq:def_muR_r}, and note that the singular case of $\int \rhat^{-1} d\rhat$ is introduced whenever $2m=n$; that is, whenever $n$ is even. Proceeding using the constants $G_n$, $B_n$ and $W_n$ from (\ref{eq:Wn_Def_text_sum}) gives
\begin{align}
\notag  &2\pi G \mu_{RL}(r;n) = \sum_{m=0}^{\infty} \frac{G_{2m}}{r^{2m+1}}\int_{a_2}^{r} -\vsqflat \left( \frac{n}{a_2} \right) \left( \frac{a_2}{\rhat} \right)^{n+1} \rhat^{2m}  d\rhat + g_{2m} r^{2m} \int_{r}^{\infty} -\vsqflat \left( \frac{n}{a_2} \right) \left( \frac{a_2}{\rhat} \right)^{n+1} \frac{1}{\rhat^{2m+1}} d\rhat\\
\notag  &= -\vsqflat \frac{n}{a_2} \Bigg[ \sum_{\begin{subarray}{c} m=0\\ 2m\ne n \end{subarray}}^{\infty} G_{2m} \frac{ a_2^{n+1}}{r^{2m+1}}  \bigg[ \frac{\rhat^{2m-n}}{2m-n}\bigg]\bigg|_{a_2}^{r} + B_n G_n\frac{a_2^{n+1}}{r^{n+1}} \log \left( \rhat  \right) \bigg|_{a_2}^{r} + \sum_{m=0}^{\infty} G_{2m} r^{2m} a_2^{n+1} \bigg[ \frac{-\rhat^{-(2m+n+1)}}{2m+n+1} \bigg|_r^\infty \bigg] \Bigg]\\
\label{eq:mu_rl_derv_num_1}  &= -\vsqflat \frac{n}{a_2} \Bigg[ \sum_{\begin{subarray}{c} m=0\\ 2m\ne n \end{subarray}}^{\infty} \frac{G_{2m}}{2m-n}  \bigg[ \left( \frac{a_2}{r} \right)^{n+1}-\left( \frac{a_2}{r} \right)^{2m+1} \bigg]  + B_nG_n \left(\frac{a_2}{r}\right) \log \left( \frac{r}{a_2}  \right) + \sum_{m=0}^{\infty} \frac{G_{2m}}{2m+n+1} \left( \frac{a_2}{r} \right)^{n+1} \Bigg]\\
\label{eq:mu_rl_derv_num_2} & = -\vsqflat \frac{n}{a_2} \Bigg[ \sum_{\begin{subarray}{c} m=0\\ 2m\ne n \end{subarray}}^{\infty} \frac{-G_{2m}}{2m-n} \left( \frac{a_2}{r} \right)^{2m+1} + \left( \frac{a_2}{r} \right)^{n+1} \bigg[ W_n + B_n G_n \log \left( \frac{r}{a_2} \right) \bigg]  \Bigg]
\end{align}
which is \eqref{eq:rho_RR_text}, as required. And so, Lemma \ref{lem:muR_series} has been proved, and Lemma \ref{lem:muL_series} may also be proven using the same methods.

Note that while \eqref{eq:mu_rl_derv_num_1} and \eqref{eq:mu_rl_derv_num_2} are formally equivalent, when the resulting series are truncated \eqref{eq:mu_rl_derv_num_1} is better behaved numerically owing to the cancellation of certain series terms.

\section{Asymptotic Behaviour of Density} \label{sec:asympt-behav-dens}

In order to prove the results of Theorem \ref{thm:cyl_mass_asym_thms_all} for disc galaxies, it is first useful to prove a Lemma relating the asymptotic
behaviour of disc density $\mu(r)$ to that of its circular velocity $\vsqr(r)$ as $r \to \infty$. Here the two profiles are related by equations \eqref{eq:1}/\eqref{eq:4} as before.

\begin{lem}\label{lem:mudisc_asym_infinity_app}
If the velocity profile $\vsqr(r)$ of a flat disc galaxy has smooth asymptotic behaviour
\begin{equation}
\label{eq:vsqr_asym_r_inf_app}
\vsqr(r) \sim C/r^n, \quad \text{as} \quad  r \to \infty
\end{equation}
for some integer $n>0$ and non-zero constant $C$, then the corresponding asymptotic behaviour of disc radial density $\mu_{2D}(r)$ is
\begin{equation}
\label{eq:rhodisc_asym_r_inf_app}
  2 \pi G \mudisc(r) \sim \begin{cases}
    \displaystyle \frac{C}{r}+\sum_{m=1}^{\infty}\frac{G_{2m}}{r^{2m+1}}\int_{0}^{R}\frac{dV^{2}(\hat{r})}{d\hat{r}}\hat{r}^{2m}d\hat{r} & ,\  n=0\\
\\
    \displaystyle \sum_{m=1}^{\infty}\frac{G_{2m}}{r^{2m+1}}\int_{0}^{R}\frac{dV^{2}(\hat{r})}{d\hat{r}}\hat{r}^{2m}d\hat{r}\\
    \displaystyle \quad +nC  \sum_{\begin{subarray}{c}
m=1\\
2m \ne n\end{subarray}}^{\infty}\frac{G_{2m}}{2m-n}\frac{R^{2m-n}}{r^{2m+1}} -\frac{nC}{r^{n+1}}\left[W_{n}+B_{n}G_{n}\ln\left(\frac{r}{R}\right)\right] & ,\ n>0
  \end{cases}
\end{equation}
as $r \to \infty$.
\begin{proof}
Choose a large but fixed radius $R$, such that for all $\rhat \ge R$, $\vsqr(\rhat) \sim C/\rhat^n$, and moreover assume that $d \vsqr(\rhat)/d\rhat \sim -nC/\rhat^{n+1}$ over the same range. Moreover, let $r$ be greater than $R$.

The integrals in \eqref{eq:4} over $\int_0^\infty$ can now be split into integrals over $\int_0^R$ and $\int_R^\infty$, and the asymptotic approximation to $d \vsqr(\rhat)/d\rhat$ applied to the upper intervals. This gives
\begin{align}
  \label{eq:mudisc_derv_step_05}  2\pi G \mudisc(r)\sim&\sum_{m=0}^{\infty}G_{2m}\int_{0}^{R}\frac{dV^{2}(\hat{r})}{d\hat{r}}\frac{r_{<}^{2m}}{r_{>}^{2m+1}}d\hat{r} +\sum_{m=0}^{\infty}G_{2m}\int_{R}^{\infty}\frac{-nC}{\hat{r}^{n+1}}\frac{r_{<}^{2m}}{r_{>}^{2m+1}}d\hat{r}  
\end{align}
as $r \to \infty$. Following this declare the terms $\mbb{T}_1$, $\mbb{T}_2$ as
\begin{align}
\label{eq:T_1_app}  \mathbb{T}_1(r)&=\sum_{m=0}^{\infty}G_{2m}\int_{0}^{R}\frac{dV^{2}(\hat{r})}{d\hat{r}}\frac{r_{<}^{2m}}{r_{>}^{2m+1}}d\hat{r}\\
\label{eq:T_2_app}  \mathbb{T}_2(r)&=\sum_{m=0}^{\infty}G_{2m}\int_{R}^{\infty}\frac{-nC}{\hat{r}^{n+1}}\frac{r_{<}^{2m}}{r_{>}^{2m+1}}d\hat{r}
\end{align}
be the series on the left and right respectively, which will now be examined in turn.

Firstly consider $\mbb{T}_1$: Since $r>R$ it follows that in $\mbb{T}_1$, the shorthands $r_<=\text{min}(r,\rhat)$ and $r_>=\text{max}(r,\rhat)$ are given by $r_<=\rhat$ and $r_>=r$. Applying these and integrating the $m=0$ term in the series gives
\begin{align}
\label{eq:term_t1_derv_05_app}
\mbb{T}_{1}(r) %=\sum_{m=0}^{\infty}\frac{G_{2m}}{r^{2m+1}}\int_{0}^{R}\frac{dV^{2}(\hat{r})}{d\hat{r}}\hat{r}^{2m}d\hat{r}\\
=\frac{G_{0}}{r}\left[V^{2}(R)-V^{2}(0)\right]+\sum_{m=1}^{\infty}\frac{G_{2m}}{r^{2m+1}}\int_{0}^{R}\frac{dV^{2}(\hat{r})}{d\hat{r}}\hat{r}^{2m}d\hat{r}
\end{align}
Now since $V^2(R) \sim C/R^n$, and $\vsqr(0)=0$, and $G_0=1$, it follows that
\begin{equation}
\label{eq:term_t1_derv_app}
  \mbb{T}_1(r)=\frac{C}{r}\frac{1}{R^{n}}+\sum_{m=1}^{\infty}\frac{G_{2m}}{r^{2m+1}}\int_{0}^{R}\frac{dV^{2}(\hat{r})}{d\hat{r}}\hat{r}^{2m}d\hat{r}
\end{equation}
Note that since $R$ is constant, the integrals do not depend on variable $r$, and so $\mathbb{T}_1=A_1/r+\sum_{n=3}^{\infty } A_n/r^n$ for constants $A_n$. Thus the first term in $\mbb{T}_1(r)$ is $O(1/r)$, with the rest being more rapidly decreasing functions of $r$.

Next consider the series $\mbb{T}_2$ in \eqref{eq:T_2_app}: To begin, the important case of $n=0$ is considered. In this case $\vsqr(r) \sim C$ and so $d\vsqr(\rhat)/d\rhat \sim 0$ when $\rhat>R$; and this gives $\mbb{T}_2(r)=0$. This exceptional case is of significance in the main text as it represents a velocity profile, and hence a rotation curve, which is asymptotically ``flat'' out to infinity.

For all other $n>0$, because $r>R$ the integrals in $\rhat$ over $\int_R^\infty$ must be split into integrals over $\int_R^r$ and $\int_r^\infty$. Over the first interval, the shorthands $r_<$ and $r_>$ become $r_{<}=\rhat$ and $r_{>}=r$, and over the second interval $r_{<}=r$ and $r_>=\rhat$.

Applying this split to \eqref{eq:T_2_app}, and integrating over $\rhat$ -- using the term $B_n$ as in appendix \ref{sect:app_elem_den_results} to indicate the exceptional integral $\int \rhat^{-1} d\rhat=\ln(\rhat)$ --  the series $\mbb{T}_2$ becomes
\begin{align*}
&\mbb{T}_{2}(r)=\sum_{m=0}^{\infty}G_{2m}\left( \int_{R}^{r}+\int_{r}^{\infty}\right)\frac{-nC}{\hat{r}^{n+1}}\frac{r_{<}^{2m}}{r_{>}^{2m+1}}d\hat{r}\\
&=-nC\left[\sum_{m=0}^{\infty}\frac{G_{2m}}{r^{2m+1}}\int_{R}^{r}\hat{r}^{2m-n-1}d\hat{r}+G_{2m}r^{2m}\int_{r}^{\infty}\frac{d\hat{r}}{\hat{r}^{2m+n+2}}\right]\\
&=-nC\left[\sum_{\begin{subarray}{c}
m=0\\
2m \ne n\end{subarray}}^{\infty}\frac{G_{2m}}{r^{2m+1}}\frac{\hat{r}^{2m-n}}{2m-n}\bigg|_{R}^{r}+\frac{B_{n}G_{n}}{r^{n+1}}\ln\left(\frac{r}{R}\right)+\sum_{m=0}^{\infty}\frac{-G_{2m}r^{2m}}{2m+n+1}\frac{1}{\hat{r}^{2m+n+1}}\bigg|_{r}^{\infty}\right]\\
&=-\frac{nC}{r^{n+1}}\left[\sum_{\begin{subarray}{c}
m=0\\
2m \ne n\end{subarray}}^{\infty}\frac{G_{2m}}{2m-n}+\sum_{m=0}^{\infty}\frac{G_{2m}}{2m+n+1}+B_{n}G_{n}\ln\left(\frac{r}{R}\right)\right] -nC\left[\sum_{\begin{subarray}{c}
m=0\\
2m \ne n\end{subarray}}^{\infty}\frac{-G_{2m}}{2m-n}\frac{R^{2m-n}}{r^{2m+1}}\right]
\end{align*}
% Upon substituting in the limits of integration, certain powers of $r^{2m}$ are seen to cancel, giving
% \begin{align*}
% \mbb{T}_2(r)
% &=-\frac{nC}{r^{n+1}}\left[\sum_{\begin{subarray}{c}
% m=0\\
% 2m \ne n\end{subarray}}^{\infty}\frac{G_{2m}}{2m-n}+\sum_{m=0}^{\infty}\frac{G_{2m}}{2m+n+1}+B_{n}G_{n}\ln\left(\frac{r}{R}\right)\right] -nC\left[\sum_{\begin{subarray}{c}
% m=0\\
% 2m \ne n\end{subarray}}^{\infty}\frac{-G_{2m}}{2m-n}\frac{R^{2m-n}}{r^{2m+1}}\right] \\
% %&\qquad -nC\left[\sum_{\begin{subarray}{c} m=0\\ 2m \ne n\end{subarray}}^{\infty}\frac{-G_{2m}}{2m-n}\frac{R^{2m-n}}{r^{2m+1}}\right]
% \end{align*}
Now replacing the series in the first bracket using the $W_n$ constants \eqref{eq:Wn_Def_text_sum}/\eqref{eq:w_n_log_def}, and separating out the $m=0$ term of the second bracket gives
\begin{equation}
\label{eq:term_t2_derv_app}
\mbb{T}_2(r)=-\frac{C}{r}\frac{1}{R^{n}}+nC\left[\sum_{\begin{subarray}{c}
m=1\\
2m \ne n\end{subarray}}^{\infty}\frac{G_{2m}}{2m-n}\frac{R^{2m-n}}{r^{2m+1}}\right]-\frac{nC}{r^{n+1}}\left[W_{n}+B_{n}G_{n}\ln\left(\frac{r}{R}\right)\right]
\end{equation}
Note that as with $\mbb{T}_1$ in \eqref{eq:term_t1_derv_app}, only the first term is $O(1/r)$, as this is the case of $n>0$, all other terms are more rapidly decreasing functions of $r$.

The results \eqref{eq:term_t1_derv_app} and \eqref{eq:term_t2_derv_app}, along with the case of $\mbb{T}_2(r)=0$ when $n=0$, can now be applied to equation \eqref{eq:mudisc_derv_step_05} to obtain \eqref{eq:rhodisc_asym_r_inf_app}, proving the lemma. Note that when $n>0$, the $O(1/r)$ terms in $\mbb{T}_1$ and $\mbb{T}_2(r)$ cancel, and $\mu(r)$ has leading order behaviour of at most $O(1/r^3)$ terms as $r \to \infty$. The exception to this is the $n=0$ case, where the term from $\mbb{T}_1$ is not cancelled and $\mu(r)$ has leading order behaviour $O(1/r)$ instead.
\end{proof}
\end{lem}

\subsection{Proof of Theorem \ref{thm:cyl_mass_asym_thms_all}}\label{sec:proof-theor-refthm:c}

To prove the results of theorem \ref{thm:cyl_mass_asym_thms_all}, simply let \textbf{I)} $n=0$, \textbf{II)} $n=1$, and \textbf{III)} $n>1$ in \eqref{eq:rhodisc_asym_r_inf_app}. Recalling from Table \ref{tab:w_n_table_text} and section \ref{sect:app_elem_den_results} that $W_1=B_1=0$, the first two of these give

\begin{align}
\text{\textbf{I)}}\quad 2 \pi G \mudisc(r) &\sim \frac{C}{r}+O(1/r^3)\\
\text{\textbf{II)}}\quad 2 \pi G \mudisc(r) &\sim\sum_{m=1}^{\infty}\frac{G_{2m}}{r^{2m+1}}\int_{0}^{R}\frac{dV^{2}(\hat{r})}{d\hat{r}}\hat{r}^{2m}d\hat{r}+C  \sum_{m=1}^{\infty}\frac{G_{2m}}{2m-1}\frac{R^{2m-1}}{r^{2m+1}}
\end{align}
And so in case \textbf{I)} $\mu(r) \sim O(1/r)$ and in \textbf{II)} $\mu(r)\sim O(1/r^3)$ as required. For case \textbf{III)}, there are differences when $n=2$ and $n>2$, but both cases lead to the same conclusions. Recalling that $B_2=1$, $B_3=W_3=0$, and integrating by parts, \eqref{eq:rhodisc_asym_r_inf_app} gives
\begin{align}
\textbf{III)} \quad n=2, \quad 2 \pi G \mudisc(r) &\sim -\frac{2CG_2}{r^{3}}\ln(r) +O(1/r^3)\\
\notag n>2, \quad 2 \pi G \mudisc(r) &\sim \frac{1}{r^3}\left(G_2\int_{0}^{R}\frac{dV^{2}(\hat{r})}{d\hat{r}}r^2  d\hat{r} +\frac{nG_2}{2-n}R^{2-n} \right) + \cdots \\
&\sim -\frac{G_2}{r^3}\left(\frac{2}{n-2}R^{2-n}   +2\int_{0}^{R} V^{2}(\hat{r}) \rhat  d\hat{r}\right) +\cdots
\end{align}
Here the ellipses denote terms more rapidly decreasing at infinity. It can be seen that the coefficients of the leading order terms in both cases are negative, and so $\mu(r)$ will eventually become negative as $r \to \infty$, as was required to be shown for case \textbf{III)}.

With no further singularities in $\mu(r)$, using \eqref{eq:Mdisc_2D_def} the remaining results of Theorem \ref{thm:cyl_mass_asym_thms_all} follow, and thus it is seen that the velocity profile $\vsqr(r)$ is restricted to be asymptotically Keplerian at infinity if the resulting disc is to be physically reasonable.

\subsection{Proof of Theorem \ref{thm:mudisc_discont_vel_text}}\label{sec:proof-theor-refthm:m}
To show the effect of a discontinuity in velocity, let the $\vsqr(r)$ have a single discontinuity at $r=a$ of magnitude $\Delta \vsqr_a = \vsqr(a+0)-\vsqr(a-0)\ne 0$. Consider the derivation of the term $\mbb{T}_1$ in lemma \ref{lem:mudisc_asym_infinity_app}. Without loss of generality, the upper bound $R$ may be chosen so that $a<R$. Integration of the $m=0$ series term in  \eqref{eq:T_1_app}/\eqref{eq:term_t1_derv_05_app} now gives instead for $\mbb{T}_1$
\begin{align*}
\mbb{T}_{1}(r)
&= \frac{G_{0}}{r}\left( \int_0^a + \int_a^R \right)  \frac{dV^{2}(\hat{r})}{d\hat{r}} d\rhat+  \sum_{m=1}^{\infty}\frac{G_{2m}}{r^{2m+1}}\int_{0}^{R}\frac{dV^{2}(\hat{r})}{d\hat{r}}\hat{r}^{2m}d\hat{r}\\
&=\frac{G_0}{r}\left[V^{2}(R)-\Delta V^{2}_a -V^{2}(0)\right]+\sum_{m=1}^{\infty}\frac{G_{2m}}{r^{2m+1}}\int_{0}^{R}\frac{dV^{2}(\hat{r})}{d\hat{r}}\hat{r}^{2m}d\hat{r}
\end{align*}
Now $V^2(R) \sim \divfrac{C}{R^n}$, $\vsqr(0)=0$, and $G_0=1$, and so the resulting $\mbb{T}_1$ is as before in \eqref{eq:term_t1_derv_app} but with an additional $-\Delta \vsqr_a/r$ term.
\begin{equation}
\label{eq:term_t1_derv_app_discont}
  \mbb{T}_1(r)=-\frac{\Delta\vsqr_a}{r} +\frac{C}{r}\frac{1}{R^{n}}+\sum_{m=1}^{\infty}\frac{G_{2m}}{r^{2m+1}}\int_{0}^{R}\frac{dV^{2}(\hat{r})}{d\hat{r}}\hat{r}^{2m}d\hat{r}
\end{equation}
Thus on adding this to $\mbb{T}_2(r)$, an overall $O(1/r)$ term will remain, even when $n>0$, and so
\begin{equation}
2 \pi G \mudisc(r) \sim -\frac{\Delta \vsqr_a}{r}+O(1/r^3)
\end{equation}
which has leading order behaviour $O(1/r)$ as $r \to \infty$ regardless of the asymptotic behaviour of $\vsqr$, which proves the theorem.

\section{Disc Mass Profiles}\label{sec:disc-mass-profiles}

In order to derive the expression of the elementary mass profiles given in lemmas \ref{lem:GML_text} and \ref{lem:GMR_text}, first certain preliminary lemmas are needed.

\subsection{Preliminary Lemmas}\label{sec:preliminary-lemmas}

The following lemmas are concerned with certain convergent series in the constants $G_{2m}$ defined in \eqref{eq:3}. It will be useful to recall the complete elliptic integrals of the first and second kind\cite{G+R}\cite{Good2001}\cite{abramowitz+stegun}\footnote{Note that many software packages follow the notation of \cite{abramowitz+stegun}, which uses $K(m)$ and $E(m)$ where $m=k^2$.}, $K(k)$ and $E(k)$, and particularly their series definitions for $0\le k <1$,
\begin{equation}
\label{eq:elliptic_integral_defs}
(2/\pi) K(k)=\sum_{m=0}^{\infty} G_{2m} k^{2m}, \qquad \text{and} \qquad (2/\pi) E(k)=\sum_{m=0}^{\infty} \frac{G_{2m}}{1-2m} k^{2m}
\end{equation}
and their special values $K(0)=E(0)=\pi/2$, $E(1)=1$, and that $K(k)$ has only a logarithmic singularity as $k\to 1$. Here Abel's theorem for the convergence of power series\cite{wittakerwatson} will also be used to take limits of expressions as $k \to 1$.

Letting $k \to 1$ in the definition of $E(k)$ and using $E(1)=1$ immediately gives the first required lemma
\begin{lem} \label{lem:G_2m_sum_res_m1}
  \begin{equation}
\label{eq:G_2m_sum_res_m1_lem} \sum_{m=0}^{\infty} \frac{G_{2m}}{2m-1}=-\frac{2}{\pi}    
  \end{equation}
\end{lem}
Following this, it is noted here without proof that disregarding constants of integration
\begin{equation}
\label{eq:g2m_sum_2mp2}
\sum_{m=0}^\infty \frac{G_{2m}}{2m+2} k^{2m}= \frac{1}{k^2}\int k K(k)  dk = \frac{2}{\pi k^2 } \left[E(k) - \left(1-k^2\right) K(k) \right]
\end{equation}
Then letting $k \to 1$ in the expression, noting that the factor $(1-k)$ cancels the logarithmic singularity of $K(k)$, gives
\begin{lem} \label{lem:G_2m_sum_res_p2}
  \begin{equation}
    \label{eq:G_2m_sum_res_p2_lem} \sum_{m=0}^{\infty} \frac{G_{2m}}{2m+2}=\frac{2}{\pi}
  \end{equation}
\end{lem}
Similarly, it can be noted that
\begin{equation}
\label{eq:g2m_sum_2mm1sq}
\sum_{m=0}^\infty \frac{G_{2m}}{\left(2m-1\right)^2} k^{2m} = \frac{2}{\pi} \left[2 E(k) - \left(1-k^2\right) K(k) \right]
\end{equation}
so again letting $k \to 1$ gives
\begin{lem} \label{lem:G_2m_sum_res_m1s}
  \begin{equation}
    \label{eq:G_2m_sum_res_m1s_lem} \sum_{m=0}^{\infty} \frac{G_{2m}}{\left(2m-1\right)^2}=\frac{4}{\pi}
  \end{equation}
\end{lem}
And in the same way, using the series
\begin{equation}
\label{eq:g2m_sum_2mp2sq}
  \sum_{m=0}^\infty \frac{G_{2m}}{\left(2m+2\right)^2} k^{2m} = \frac{2}{\pi k^2} \left[2 E(k) - \left(1-k^2\right) K(k) -\frac{\pi}{2}\right],
\end{equation}
\begin{lem} \label{lem:G_2m_sum_res_p2s}
  \begin{equation}
    \label{eq:G_2m_sum_res_p2s_lem} \sum_{m=0}^{\infty} \frac{G_{2m}}{\left(2m+2\right)^2}=\frac{4}{\pi}-1
  \end{equation}
\end{lem}

\subsection{Elementary mass profile results} \label{sec:elem-mass-prof}

A method for deriving the elementary mass profile series is now described. For brevity, only the series for $M_R(r;n)$ in lemma \ref{lem:GMR_text} will be described explicitly. The procedure is similar for $M_L(r;n)$ in \ref{lem:GML_text}, with no exceptional cases.

Turning to the piecewise expression \eqref{eq:GM_R_text} for $M_R$, consider the \emph{indefinite} integrals
 \begin{align}
\label{eq:GMRL_app}  GM_{RL}(r;n)=& \int r \cdot 2\pi G\mu_{RL}(r;n)dr\\
\label{eq:GMRR_app}  GM_{RR}(r;n)=& \int r \cdot 2\pi G\mu_{RR}(r;n)dr
 \end{align}
Leaving aside constants of integration, it will be shown that these are the functions \eqref{eq:GMRL_text} and \eqref{eq:GMRR_text} being sought. Using \eqref{eq:rho_RL_text} in \eqref{eq:GMRL_app}, and leaving aside formal constants of integration gives
\begin{align}
\notag G M_{RL}(r;n)&= -\vsqflat \frac{n}{a_{2}}\sum_{m=0}^{\infty}\frac{G_{2m}}{2m+n+1}\int r\left(\frac{r}{a_{2}}\right)^{2m}dr\\
\label{eq:GMRL_app_derved} & =  - \vsqflat  na_{2}\sum_{m=0}^{\infty}\frac{G_{2m}}{(2m+n+1)(2m+2)}\left(\frac{r}{a_{2}}\right)^{2m+2}
\end{align}
Which is the desired result \eqref{eq:GMRL_text}. In the same way, using \eqref{eq:rho_RR_text} in \eqref{eq:GMRR_app} for the case of $n \ne 1$ gives
\begin{align}
\notag G M_{RR}(r;n)= & \vsqflat  \frac{n}{a_{2}}\left[\sum_{\begin{subarray}{c}
m=0\\
2m \ne n\end{subarray}}^{\infty}\frac{G_{2m}}{2m-n}\int r\left(\frac{a_{2}}{r}\right)^{2m}dr+B_{n}G_{n}\int r\left(\frac{a_{2}}{r}\right)^{n+1}\ln\left(\frac{a_{2}}{r}\right)dr-W_{n}\int r\left(\frac{a_{2}}{r}\right)^{n+1}dr\right]\\
\label{eq:GMRR_app_derved} = &  \vsqflat  n a_{2}\left[\sum_{\begin{subarray}{c}
m=0\\
2m \ne n \end{subarray}}^{\infty}\frac{-G_{2m}}{(2m-n)\left(2m-1\right)}\left(\frac{a_{2}}{r}\right)^{2m-1}+\left(\frac{a_{2}}{r}\right)^{n-1}\left[\frac{W_{n}}{n-1}+\frac{B_{n}G_{n}}{n-1}\left[\frac{1}{n-1}-\ln\left(\frac{a_{2}}{r}\right)\right]\right]\right]
\end{align}
Which is the $n \ne 1$ case of \eqref{eq:GMRR_text}. Moreover, recalling that $B_1=W_1=0$, terms containing these need only be omitted in this result to obtain the $n=1$ case in \eqref{eq:GMRR_text} as well. And so the indefinite integrals \eqref{eq:GMRL_app}/\eqref{eq:GMRR_app} are the functions \eqref{eq:GMRL_text}/\eqref{eq:GMRR_text} being sought.

It remains to show $GM_R(r;n)$, which is defined by \eqref{eq:elemen_scaled_mass_profiles}, is given by \eqref{eq:GM_R_text}. Formally, using \eqref{eq:rho_R_piecewise_text} in \eqref{eq:elemen_scaled_mass_profiles} gives
\begin{equation}
  \label{eq:GM_R_piecewise_full_appendix}
  GM_R(r;n)=\begin{cases}
    GM_{RL}(r;n) - GM_{RL}(0;n)&, r<a_2\\
    GM_{RR}(r;n) - GM_{RR}(a_2;n) \\
    \quad + \left(GM_{RL}(a_2;n) - GM_{RL}(0;n)\right) &, r>a_2
  \end{cases}
\end{equation}
This representation can be simplified to \eqref{eq:GM_R_text} by finding the values of $GM_{RL}(0;n)$, $GM_{RL}(a_2;n)$, and $GM_{RR}(a_2;n)$.
  
By setting $r=0$ in \eqref{eq:GMRL_app_derved}/\eqref{eq:GMRL_text}, it can immediately be seen that $GM_{RL}(0;n)=0$. For the values of $GM_{RL}(a_2;n)$ and $GM_{RR}(a_2;n)$, first consider the case of $n \ne 1$. Letting $r=a_2$ in \eqref{eq:GMRL_app_derved} and using partial fractions on quadratic denominators gives
\begin{align*}
G M_{RL}(a_{2};n)= & - \vsqflat  na_{2}\sum_{m=0}^{\infty}\frac{G_{2m}}{(2m+n+1)(2m+2)}=  \vsqflat  \frac{ na_{2}}{n-1}\sum_{m=0}^{\infty}\frac{G_{2m}}{2m+n+1}-\frac{G_{2m}}{2m+2}\end{align*}
Applying this same process to \eqref{eq:GMRR_app_derved}/\eqref{eq:GMRR_text}, and also expanding $W_n$ using its definition in \eqref{eq:Wn_Def_text_sum} gives
\begin{align*}
&G M_{RR}(a_{2};n)= \vsqflat  n a_2 \left[\sum_{\begin{subarray}{c}
m=0\\
m\ne\frac{n}{2}\end{subarray}}^{\infty}\frac{-G_{2m}}{(2m-n)\left(2m-1\right)}+\frac{W_{n}}{n-1}+\frac{B_{n}G_{n}}{(n-1)^{2}}\right]\\
= & \vsqflat n a_2 \left[\sum_{\begin{subarray}{c}
m=0\\
m\ne\frac{n}{2}\end{subarray}}^{\infty}\frac{1}{n-1}\left(\frac{G_{2m}}{2m-1}-\frac{G_{2m}}{2m-n}\right)+\frac{1}{n-1}\left[\sum_{m=0}^{\infty}\frac{G_{2m}}{(2m+n+1)}+\sum_{\begin{subarray}{c}
m=0\\
m\ne\frac{n}{2}\end{subarray}}^{\infty}\frac{G_{2m}}{a_{2}(2m-n)}\right]+\frac{B_{n}G_{n}}{(n-1)^{2}}\right]\\
= & \vsqflat  \frac{ n a_1 }{n-1}\sum_{m=0}^{\infty}\frac{G_{2m}}{2m+n+1}+\frac{G_{2m}}{2m-1}\end{align*}
Subtracting these two results, and using lemmas \ref{lem:G_2m_sum_res_m1} and \ref{lem:G_2m_sum_res_p2} then gives
\begin{align}
\notag  G M_{RR}(a_{2};n)-G M_{RL}(a_{2};n)  &=\vsqflat \frac{na_{2}}{n-1}\sum_{m=0}^{\infty}\frac{G_{2m}}{2m-1}+\frac{G_{2m}}{2m+2}\\
\label{eq:MRR_MRL_n_ne_1}&=\vsqflat \frac{na_{2}}{n-1}\left(-\frac{2}{\pi}+\frac{2}{\pi}\right)=0
\end{align}
And so $GM_{RL}(a_2;n)$ and $G_{RR}(a_2;n)$ are equal when $n \ne 1$. However, this result does not hold for the special case\footnote{The case of $n=0$ is not considered formally in this paper as it leads to disc with infinite mass.} of $n=1$, as the factors $2m+2$ and $2m-1$ are repeated in their respective denominators, so that the fractions are already in decomposed form. Instead when $n=1$, the formulas \eqref{eq:GMRL_text} and \eqref{eq:GMRR_text} for $r=a_2$ must be evaluated using the results of lemmas \ref{lem:G_2m_sum_res_m1s} and \ref{lem:G_2m_sum_res_p2s}. When applied these give
\begin{align*}
G M_{RL}(a_{2};1)&=- \vsqflat  a_{2}\sum_{m=0}^{\infty}\frac{G_{2m}}{(2m+2)^{2}}=-\vsqflat a_{2}\left(\frac{4}{\pi}-1\right)\\
G M_{RR}\left(a_{2};1\right)&=  \vsqflat a_{2}\left[\sum_{m=0}^{\infty}\frac{-G_{2m}}{\left(2m-1\right)^{2}}+0+0\right]=\vsqflat  a_{2}\left(-\frac{4}{\pi}\right)
\end{align*}
Subtraction these values then gives the result
\begin{align}
\label{eq:MRR_MRL_n_eq_1} GM_{RL}\left(a_{2};1\right)-&GM_{RR}(a_{2};1)=\vsqflat a_{2}\left(-\frac{4}{\pi}+1 +\frac{4}{\pi}\right) =\vsqflat a_{2}
\end{align}
So combining \eqref{eq:MRR_MRL_n_eq_1} and \eqref{eq:MRR_MRL_n_ne_1} gives for all $n$,
\begin{align*}
G M_{RL}\left(a_{2};n\right)-G M_{RR}(a_{2};n)&=  \vsqflat a_2 \delta_{1n} = \begin{cases}
\vsqflat  a_{2} & ,n=1\\
0 & ,n\ne1\end{cases}
\end{align*}
where $\delta_{1n}$ is the Kronecker delta. Applying this and $GM_{RL}(0;n)=0$ to \eqref{eq:GM_R_piecewise_full_appendix} then gives the result \eqref{eq:GM_R_text} which proves lemma \ref{lem:GMR_text}. Lemma \ref{lem:GML_text} is proven similarly, and with less work as it does not possess any special cases.

\subsection{Effect of velocity perturbations on mass convergence}\label{sec:effect-veloc-pert}

First an integral directly relating $M(r)$ to general $\vsqr(r)$ is needed. Substituting (\ref{eq:1}) into (\ref{eq:Mdisc_2D_def}) and integrating, it follows that (See \cite{Nordsieck1973} eq (5))
\begin{equation}
  \label{eq:5}
  M(r) = \frac{r}{G}\int_0^\infty \vsqr(\rhat) \frac{2}{\pi r_<} \left[ K\left(\frac{r_<}{r_>}\right) - E\left(\frac{r_<}{r_>}\right) \right] d \rhat
\end{equation}
where $r_<$ and $r_>$ are as in \eqref{eq:rmin_rmax_shorthands_def}. Now, consider a perturbation $\Delta \vsqr(\rhat)$ of the velocity profile over some finite interval $p_1 < \rhat <p_2$. On this interval, the perturbation $\Delta \vsqr$ is a finite continuous function which must be zero at the endpoints $p_1,p_2$ so that the resulting galaxy has finite mass by theorem \ref{thm:mudisc_discont_vel_text}. By the linearity of \eqref{eq:5} in $\vsqr$, it follows that the corresponding perturbation in the mass profile is given by
\begin{equation}
  \label{eq:6}
  \Delta M(r) = \frac{1}{G}\int_{p_1}^{p_2} \Delta \vsqr(\rhat) \frac{2}{\pi}\frac{r}{r_<} \left[ K\left(\frac{r_<}{r_>}\right) - E\left(\frac{r_<}{r_>}\right) \right] d \rhat
\end{equation}
Note that since $\Delta \vsqr$ is non-zero only over a finite region, by \eqref{eq:GMdisctot_kep_coeff_result} it follows that $\Delta M(r) \to 0$ as $r \to \infty$, and no overall total mass is added.

However the rate of asymptotic convergence of the mass profile is affected. To examine the change in rate, assume that $r > p_2$, so that $r_< = \rhat$ and $r_>=r$ over the finite interval in \eqref{eq:6}. Then, using the series in \eqref{eq:elliptic_integral_defs}, it follows that
\begin{align}
  \label{eq:delta_mass_derv}
  \Delta M(r) & = \frac{1}{G} \int_{p_1}^{p_2} \Delta \vsqr(\rhat) \sum_{m=1}^{\infty} \frac{2m}{2m-1} G_{2m} \left( \frac{\rhat}{r} \right)^{2m-1} d\rhat\\
  &= \frac{1}{G}\sum_{m=1}^{\infty} \frac{2m}{2m-1} \frac{G_{2m}}{r^{2m-1}} \int_{p_1}^{p_2} \Delta \vsqr(\rhat) \rhat^{2m-1} d\rhat
\end{align}
As $p_2 < r$ and $\Delta \vsqr$ is finite, this result is a convergent geometric series. Evaluating the $m=1$ term in the series using \eqref{eq:3} gives the change in the $O(1/r)$ asymptotic convergence rate of the mass profile as
\begin{equation}
  \label{eq:7}
  \Delta M(r) = \frac{1}{2 \pi G r} \left[ 2 \pi \int_{p_1}^{p_2} \frac{1}{2} \Delta \vsqr(\rhat) \rhat d \rhat \right] + O \left( \frac{1}{r^3} \right)
\end{equation}
The integral in the bracket can be regarded as the perturbation in the total specific kinetic energy due to $\Delta \vsqr$. Its value divided by $2 \pi G$ gives the change in the coefficient of the $O(1/r)$ convergence rate of $M(r)$ in the outer regions of the galaxy. Thus it can be seen that an overall increase in $\vsqr$ will lead to a more rapidly converging mass profile, and an overall decrease in $\vsqr$ will tend to shift greater percentages of the galaxies mass to outer regions. See for example models B,F,G and H in sections \ref{sec:example-models} and \ref{sec:example-models-1}.

\end{document}